\documentclass{LMCS}

\def\doi{8 (1:21) 2012}
\lmcsheading%
{\doi}
{1--26}
{}
{}
{Nov.~\phantom09, 2010}
{Mar.~\phantom08, 2012}
{}

\usepackage{enumerate,hyperref}
\usepackage{amsfonts}
\usepackage{amssymb}
\usepackage{amsmath}

\usepackage{natbib}
\let\cite=\citep   
\let\citeyear=\citeyearpar

\newenvironment{prog}{\begin{array}[t]{@{}l@{}}}{\end{array}}

\usepackage{amsthm}

\newtheoremstyle{theorem}{\topsep}{\topsep}%
     {\itshape}%
     {}%
     {\bfseries}%
     {.}%
     {10pt}%
     {\thmname{#1}\thmnumber{ #2}\thmnote{ (#3)}}%

\theoremstyle{theorem}
\newtheorem{theorem}{Theorem}[section]

\newtheorem{proposition}[theorem]{Proposition}

\newtheorem{EXAMPLE}[theorem]{Example}

\renewcommand{\qedsymbol}{\bbox}
\newcommand{\bbox}{\vrule height7pt width4pt depth1pt}

\newcommand{\sat}{\models}
\newcommand{\rimp}{\Rightarrow}

\newcommand{\<}{\langle}
\renewcommand{\>}{\rangle}

\renewcommand{\phi}{\varphi}

\newcommand{\K}{\mathcal{K}}
\newcommand{\I}{\mathcal{I}}
\newcommand{\R}{\mathcal{R}}
\newcommand{\A}{\mathtt{A}}

\newcommand{\Cc}{\mathcal{C}}
\newcommand{\cP}{\mathcal{P}}
\newcommand{\cM}{\mathcal{M}}
\newcommand{\cK}{\mathcal{K}}

\newcommand{\sent}{\mathsf{send}}
\newcommand{\recv}{\mathsf{recv}}
\newcommand{\intercept}{\mathsf{intercept}}
\newcommand{\sees}{\mathsf{has}}
\newcommand{\tspace}{~~~~~}

\newcommand{\sL}{\ell}
\newcommand{\btrue}{\mbox{\textbf{true}}}
\newcommand{\bfalse}{\mbox{\textbf{false}}}

\newcommand{\ADYi}{\A^{\scriptscriptstyle\rm DY}_i}
\newcommand{\ADY}{\A^{\scriptscriptstyle\rm DY}}

\newcommand{\ADDGi}{\A^{\scriptscriptstyle\rm DDG}_i}
\newcommand{\ALi}{\A^{\scriptscriptstyle\rm L}_i}

\newcommand{\encr}[2]{\{\!\hspace{-.7pt}|#1|\!\hspace{-.7pt}\}_{#2}}
\newcommand{\inv}[1]{#1^{-1}}

\newcommand{\union}{\ensuremath{\cup}}

\newcommand{\derivesDY}{\vdash_{\scriptscriptstyle DY}}
\newcommand{\derivesL}{\vdash_{\scriptscriptstyle L}}

\newcommand{\YES}{\mbox{``Yes''}}
\newcommand{\NO}{\mbox{``No''}}
\newcommand{\DONTK}{\mbox{``?''}}

\newcommand{\PhiA}{\Phi_0^{\scriptscriptstyle\rm S}}

\newcommand{\LKXn}{\mathcal{L}^{\scriptscriptstyle\rm KX}_n}

\newcommand{\msg}{\mathsf{m}}
\newcommand{\ky}{\mathsf{k}}
\newcommand{\pl}{\mathsf{p}}

\newcommand{\Rule}[2]{          %
  \begin{array}{c}
  #1 \\\hline
  #2
  \end{array}}

\newcommand{\commentout}[1]{}

\newcommand{\etal}{et al.}

\newcommand{\MSG}{\cM}
\newenvironment{axiomlist}
   {\begin{wideitemize}{8ex}}
   {\end{wideitemize}}
\newenvironment{wideitemize}[1]
   {\begin{list}{$\bullet$}
                     {\setlength{\labelwidth}{#1}
                      \setlength{\leftmargin}{#1}}}
   {\end{list}}

\begin{document}

\title[Modeling Adversaries]{Modeling Adversaries in a Logic for Security Protocol Analysis}

\author[J.~Y.~Halpern]{Joseph Y.~Halpern\rsuper a}	%
\address{{\lsuper a}Cornell University\\
Ithaca, NY 14853}	%
\email{halpern@cs.cornell.edu}  %

\author[R.~Pucella]{Riccardo Pucella\rsuper b}	%
\address{{\lsuper b}Northeastern University\\
Boston, MA 02115}
\email{riccardo@ccs.neu.edu}  %
\thanks{This work was done while the second author was at Cornell University.}	%


\begin{abstract}
Logics for security protocol analysis require the formalization of an 
adversary model that specifies the capabilities of adversaries. A common
model is the Dolev-Yao model, which considers only adversaries
that can compose and replay messages, and decipher them with known
keys.  The Dolev-Yao model is a useful abstraction, but it suffers
from some drawbacks: it cannot handle the adversary knowing
protocol-specific information, and it cannot handle probabilistic
notions, such as the adversary attempting to guess the keys. We show
how we can analyze security protocols under different adversary models
by using a logic with a notion of algorithmic knowledge.  Roughly
speaking, adversaries are assumed to use algorithms to compute their
knowledge; adversary capabilities are captured by suitable restrictions
on the algorithms used. We show how we can model the standard
Dolev-Yao adversary in this setting, and how we can capture more
general capabilities including protocol-specific knowledge and
guesses. 
\end{abstract}

\maketitle

\section{Introduction}

Many formal methods for the analysis of security protocols rely on
specialized logics to rigorously prove properties of
the protocols they study.%
\footnote{Here, we take a very general view of logic, to encompass formal
methods where the specification language is implicit, or where the
properties to be checked are fixed, such as Casper
\cite{r:lowe98}, Cryptyc \cite{r:gordon03}, or the NRL Protocol Analyzer
\cite{r:meadows96}.} 
Those logics provide constructs for
expressing the basic notions involved in security protocols, such as
secrecy, recency, and message composition, as well as providing means
(either implicitly or explicitly)
for describing the evolution of the knowledge or belief of the
principals as the protocol progresses. Every such logic aims at
proving security in the presence of hostile adversaries.  To analyze the
effect of adversaries, a security logic specifies 
(again, either implicitly or
explicitly) an \emph{adversary model}, that is, a description of the
capabilities of adversaries.  Almost all existing logics are based on a
Dolev-Yao adversary model \cite{r:dolev83}. Succinctly, a Dolev-Yao
adversary can compose messages, replay them, or decipher them if she
knows the right keys, but cannot otherwise ``crack'' encrypted
messages.

The Dolev-Yao adversary is a useful abstraction, in that it
allows reasoning about protocols without worrying about the actual
encryption scheme being used.  It also has the advantage of being
restricted enough that interesting theorems can be proved with respect
to security.  However, in many ways, the Dolev-Yao model is too
restrictive. For example, it does not consider the information an
adversary may infer from properties of messages and knowledge about
the protocol that is being used. To give an extreme example, consider
what we will call the \emph{Duck-Duck-Goose} protocol:~an agent has an
$n$-bit key and, according to her protocol, sends the bits that make
up its key one by one. Of course, after intercepting these messages,
an adversary will know the key. However, there is no way for security
logics based on a Dolev-Yao adversary to argue that, at this
point, the adversary knows the key. Another limitation of the
Dolev-Yao adversary is that it does not easily capture
probabilistic arguments. After all, the adversary can always be lucky
and just \emph{guess} the appropriate key to use, irrespective of the
strength of the encryption scheme.
The importance of being able to reason about adversaries with
capabilities beyond those of a Dolev-Yao adversary is made clear when
we look at the subtle interactions between
the cryptographic protocol and the encryption scheme.  It is known that
various protocols that are secure with respect to a Dolev-Yao adversary can
be broken when implemented using encryption schemes with specific properties
\cite{r:moore88}, such as encryption systems with encryption cycles
\cite{r:abadi02a} and ones that use exclusive-or \cite{r:ryan98}.
A more refined logic for reasoning about security protocols will have to
be able to handle adversaries more general than the Dolev-Yao
adversary.

Because they effectively build in the adversary model,
many formal methods for analyzing protocols are not able to reason
directly about the effect of running a protocol against adversaries
with properties other than those built in.  
Some formal methods allow much flexibility in their description of
adversaries---for example, Casper~\cite{r:lowe98}, AVISPA~\cite{r:vigano05},
ProVerif~\cite{r:abadi05}---but they are still bound to the restrictions
of the models underlying their respective analysis methods,
generally expressed in terms of an equational theory. 
The problem is even worse
when it is not clear exactly what assumptions are implicitly being made
about the adversary.
One obvious assumption that needs to be made clear is whether the
adversary is an insider in the system or an
outsider. Lowe's \citeyear{r:lowe95} well-known man-in-the-middle attack 
against the Needham-Schroeder \citeyear{r:needham78} protocol
highlights this issue. Until then, the Needham-Schroeder
protocol had been analyzed under the assumption that the adversary had
complete control of the network, and could inject intercept and inject
arbitrary messages (up to the Dolev-Yao capabilities) into the
protocol runs. However, the adversary was always assumed to be an
outsider, not being able to directly interact with the protocol
principals as himself. Lowe showed that if the adversary is allowed to
be an \emph{insider} of the system, that is, appear to the other
principals as a bona fide protocol participant, then the
Needham-Schroeder protocol does not guarantee the authentication
properties it is meant to guarantee.

In this paper, we introduce a logic for reasoning about security
protocols that allows us to model adversaries explicitly
and naturally.
The idea is to model the adversary in terms of what the adversary
knows. 
This approach has some significant advantages.  Logics of knowledge
\cite{r:fagin95} have been shown to provide powerful methods for reasoning
about trace-based executions of protocols. 
They can be given semantics that is tied directly to protocol execution,
thus avoiding problems of having to analyze an idealized form of the
protocol, as is required, for example, in BAN logic \cite{r:burrows90}.
A straightforward application of logics of knowledge allows us to 
conclude that in the Duck-Duck-Goose protocol, the adversary knows the
key. Logics of knowledge can also be extended with probabilities
\cite{FH3,HT} so as to be able to deal with probabilistic phenomena.
Unfortunately, traditional logics of knowledge suffer from a
well-known problem known as the \emph{logical omniscience} problem: an
agent knows all tautologies and all the logical consequences of her
knowledge.  The reasoning that allows an agent to infer properties of
the protocol also allows an attacker to deduce properties that cannot
be computed by realistic attackers in any reasonable amount of time.

To avoid the logical omniscience problem, 
we use the notion of \emph{algorithmic knowledge}
\cite[Chapter 10 and 11]{r:fagin95}. 
Roughly speaking, we assume that agents
(including adversaries) have ``knowledge 
algorithms'' that they use to compute what they know.  The
capabilities of the adversary are captured by its algorithm. Hence,
Dolev-Yao capabilities can be provided by using a knowledge algorithm
that can only compose messages or attempt to decipher using known keys. 
By changing the algorithm, we can extend the capabilities of the
adversary so that it can attempt to crack the encryption scheme by factoring
(in the case of RSA), using differential cryptanalysis (in the case of
DES), or just by guessing keys,
along the lines of a model due to Lowe \citeyear{r:lowe02}.  
Moreover, our framework can also
handle the case of a principal sending the bits of its key, by
providing the adversary's algorithm with a way to check whether this is
indeed what is happening. By explicitly using algorithms, we can
therefore analyze the effect of bounding the resources of the
adversary, and thus make progress toward bridging the gap between the
analysis of cryptographic protocols and more computational accounts of
cryptography.  (See \cite{r:abadi02a} and the references therein for a
discussion on work bridging this gap.)
Note that we need both traditional knowledge and algorithmic  knowledge
in our analysis.  Traditional knowledge is used to model a principal's
beliefs about what can happen in the protocol; algorithmic knowledge is
used to model the adversary's computational limitations (for example,
the fact that it cannot factor).

The focus of this work is on developing a general and expressive
framework for modeling and reasoning about security protocols, in
which a wide class of adversaries can be represented naturally. 
In particular, we hope that the logic can provide a foundation for
comparison and evaluation of formal methods that use different
representation of adversaries.  
Therefore, we emphasize the expressiveness and representability
aspects of the framework, rather than studying the
kind of security properties that are useful in such a setting or
developing techniques for proving that properties hold in the
framework. These are all relevant questions that need to be pursued
once the framework proves useful as a specification language. 

The rest of the paper is organized as follows.  In
Section~\ref{s:protocols}, we define our model for protocol analysis
based  on the well-understood multiagent system framework, and in 
Section~\ref{s:logic} we present a logic for reasoning about implicit
and explicit knowledge.
In Section~\ref{s:adversaries}, we show how to model different
adversaries from the literature. 
In Section~\ref{s:passive}, these adversaries are passive, in that
they eavesdrop on the 
communication but do not attempt to interact with the principals of
the system; in Section~\ref{s:active}, we show how the framework can
accommodate adversaries that actively
interact with the principals by intercepting, forwarding, and
replacing messages.  
We discuss related work in Section~\ref{s:related}.

\section{Modeling Security Protocols}\label{s:protocols}

In this section, we review the multiagent system
framework of Fagin \etal{} \citeyear[Chapters 4 and 5]{r:fagin95}, and
show it can be tailored to represent security protocols.

\commentout{
The intuitive idea behind the classical
approach to knowledge and belief \cite{Hi1} is that, besides the true
state of affairs, there are a number of other possible states of
affairs, or possible worlds. Some of these possible worlds may be
indistinguishable to an agent from the true world. An agent is then
said to know a fact $\phi$ if $\phi$ is true in all the worlds he
thinks possible. The basic possible-worlds framework does not
lend itself so well to protocol analysis. Where are the possible
worlds coming from? The multiagent systems model \cite[Chapters 4 and 
5]{Fagin95} provides an answer that has
the advantage of also providing a discipline for modeling executions
of protocols.
}

\subsection{Multiagent Systems}
The multiagent systems framework  
provides a model for knowledge that has
the advantage of also providing a discipline for modeling executions
of protocols.  
A multiagent system consists of $n$ agents, each of which is in some 
local state at a given point in time. 
We assume that an agent's local
state encapsulates all the information to which the agent has
access. In the security setting, the local state of an agent might
include some initial information regarding keys, the messages she has
sent and received, and perhaps the reading of a clock. In a poker
game, a player's local state might consist of the cards he currently
holds, the bets made by other players, any other cards he has seen,
and any information he may have about the strategies of the other
players (for example, Bob may know that Alice likes to bluff, while
Charlie tends to bet conservatively). The basic framework makes no
assumptions about the precise nature of the local state.

We can then view the whole system as being in some global state, which
is a tuple consisting of each agent's local state, together with the
state of the environment, where the environment consists of everything
that is relevant to the system that is not contained in the state of
the agents. Thus, a global state has the form $(s_e, s_1,\ldots, s_n)$,
where $s_e$ is the state of the environment and $s_i$ is agent $i$'s state,
for $i = 1, \ldots , n$. The actual form of the agents' local states and
the environment's state depends on the application.

A system is not a static entity. To capture its dynamic aspects, we
define a run to be a function from time to global states. Intuitively,
a run is a complete description of what happens over time in one
possible execution of the system. A \emph{point} is a pair $(r, m)$ consisting
of a run $r$ and a time $m$. For simplicity, we take time to range over
the natural numbers in the remainder of this discussion. At a point
$(r, m)$, the system is in some global state $r(m)$. If $r(m) = (s_e, s_1,
\ldots , s_n)$, then we take $r_i(m)$ to be $s_i$, agent $i$'s local state at the
point $(r, m)$. We formally define a system 
$\R$ 
to consist of a set of runs (or
executions). It is relatively straightforward to model security
protocols as systems. Note that the adversary in a security protocol
can be modeled as just another agent. The adversary's information at a
point in a run can be modeled by his local state.

\subsection{Specializing to Security}

The multiagent systems framework is quite general.
We have a particular application in mind, namely reasoning
about security protocols, especially authentication protocols. We now
specialize the framework in a way appropriate for reasoning about
security protocols.

Since the vast majority of security protocols studied in the
literature are message-based, a natural class 
of multiagent systems to consider is that of \emph{message passing systems}
\cite{r:fagin95}. Let $\MSG$ be a fixed set of messages.  A
\emph{history} for agent $i$ (over $\MSG$) 
is a sequence of elements of the form $\sent(j,\msg)$ and $\recv(\msg)$,
where~$\msg \in \MSG$.  We think of $\sent(j,\msg)$ as representing the
event ``message $\msg$ is sent to $j$'' and $\recv(\msg)$ as representing the
event ``message~$\msg$ is received.''. 
(We also allow events corresponding to internal actions; for the
purpose of this paper, the internal actions we care about
concern adversaries eavesdropping or intercepting messages---we return
to those in Section~\ref{s:adversaries}.)
Intuitively, $i$'s history at $(r,m)$ consists of~$i$'s initial state,
which we take to be the empty sequence, followed by the sequence
describing~$i$'s actions up to time~$m$.  If~$i$ performs no
actions in round~$m$, then her history at $(r,m)$ is the same as her
history at $(r,m-1)$.  In such a message-passing system, we speak of
$\sent(j,\msg)$ and $\recv(\msg)$ as {\em events}.  For an agent $i$, let
$r_i(m)$ be agent $i$'s history in $(r,m)$.  We say that an event $e$
\emph{occurs in $i$'s history in round $m+1$ of run~$r$} if $e$ is in
(the sequence) $r_i(m+1)$ but not in $r_i(m)$.

In a message-passing system, the agent's local state at any point is
her history.  Of course, if~$h$ is the history of agent~$i$ at the
point $(r,m)$, then we want it to be the case that~$h$ describes what
happened in~$r$ up to time~$m$ from~$i$'s point of view.  To do this,
we need to impose some consistency conditions on global states.  In
particular, we want to ensure that message histories do not shrink
over time, and that every message received in round~$m$ corresponds to
a message that was sent at some earlier round.

Given a set $\MSG$ of messages, we define a \emph{message-passing
system} (over $\MSG$) to be a system 
satisfying the following constraints at all points $(r,m)$ for each
agent $i$:
\begin{axiomlist}
\item[MP1.] $r_i(m)$ is a history over $\MSG$.
\item[MP2.]
For every event $\recv(\msg)$ in $r_i(m)$ there exists a
corresponding event $\sent(i,\msg)$ in $r_j(m)$.
\item[MP3.]
$r_i(0)$ is the empty sequence
and $r_i(m+1)$ is either identical to $r_i(m)$ or the result of
appending one event to $r_i(m)$.
\end{axiomlist}
MP1 says that an agent's local state is her history, MP2 guarantees
that every message received at round~$m$ corresponds to one that was
sent earlier, and MP3 guarantees that histories do not shrink.
We note that there is no guarantee that messages are not delivered
twice, or that they are delivered at all.

A \emph{security system} is a message passing system where the message
space has a structure suitable for the interpretation of security
protocols. 
Therefore, a
security system assumes a set $\cP$ of plaintexts, as well as a set
$\cK$ of keys.  
An
encryption scheme $\Cc$ over $\cP$ and $\cK$ is the closure $\cM$ of
$\cP$ and $\cK$ under 
a key inverse operation $\mathit{inv} : \cK\rightarrow\cK$, 
a concatenation operation $\mathit{conc} :
\cM\times\cM\rightarrow\cM$, decomposition operators
$\mathit{first}:\cM\rightarrow\cM$ and
$\mathit{second}:\cM\rightarrow\cM$, an encryption operation $\mathit{encr} :
\cM\times\cK\rightarrow\cM$, and a decryption operation
$\mathit{decr}:\cM\times\cK\rightarrow\cM$, subject to the
constraints:
\begin{align*}
\mathit{first}(\mathit{conc}(\msg_1,\msg_2)) & = \msg_1\\
\mathit{second}(\mathit{conc}(\msg_1,\msg_2)) & = \msg_2\\
\mathit{decr}(\mathit{encr}(\msg,\ky),\mathit{inv}(\ky)) & = \msg.
\end{align*}
In other words, decrypting an encrypted message with the inverse of
the key used to encrypt the message yields the original message. (For
simplicity, we restrict ourselves to nonprobabilistic encryption schemes in
this paper.)  We often write 
$\inv{\ky}$ for $\mathit{inv}(\ky)$, 
$\msg_1\cdot\msg_2$ for
$\mathit{conc}(\msg_1,\msg_2)$, and $\encr{\msg}{\ky}$ for
$\mathit{encr}(\msg,\ky)$.  There is no difficulty in adding more
operations to the encryption schemes, for instance, to model hashes,
signatures, the ability to take the exclusive-or of two terms, or the
ability to compose two keys together to create a new key. We 
make no assumption in the general case as to the properties of
encryption. Thus, for instance, most concrete encryption schemes allow
collisions, that is, $\encr{m_1}{k_1}=\encr{m_2}{k_2}$ without
$m_1=m_2$ and $k_1=k_2$. (In contrast, most security protocol 
analyses assume that there are no properties of encryption schemes
beyond those specified above; this is part of the Dolev-Yao adversary
model, which we examine in more detail in Section~\ref{s:dy}.)

\newcommand{\sub}{\preccurlyeq}

Define $\sub$ on $\cM$ as the smallest relation
satisfying the following constraints: 
\begin{enumerate}[(1)]
\item $\msg\sub \msg$
\item if $\msg\sub \msg_1$, then $\msg\sub \msg_1\cdot \msg_2$
\item if $\msg\sub \msg_2$, then $\msg\sub \msg_1\cdot \msg_2$
\item if $\msg\sub \msg_1$, then $\msg\sub \encr{\msg_1}{\ky}$.
\end{enumerate}
Intuitively, $\msg_1\sub \msg_2$ if $\msg_1$ \emph{could} be used in the
construction of $\msg_2$, or, equivalently, if $\msg_1$ can be
considered part of message $\msg_2$, under any of the possible ways of
decomposing $\msg_2$. 
For example, if
$\msg=\encr{\msg_1}{\ky}=\encr{\msg_2}{\ky}$, then both
$\msg_1\sub \msg$ and $\msg_2\sub \msg$.  
If we want to establish that $\msg_1\sub \msg_2$ for
a given $\msg_1$ and $\msg_2$, then we have to recursively look at all
the possible ways in which $\msg_2$ can constructed to see if $\msg_1$
can possibly be used to construct $\msg_2$. 
Clearly, if encryption does not result in collisions, there is a
single way in which $\msg_2$ can be taken apart. 
If the cryptosystem under consideration supports others operations
(e.g., an exclusive-or operation, or operations that to combine keys 
to create new keys), additional constraints must be added to the
$\sub$ relation to account for those operations.

To analyze a particular security protocol, we first derive the
multiagent system corresponding to the protocol, 
using the approach of Fagin \etal{} \citeyear[Chapter 5]{r:fagin95}. 
Intuitively, this
multiagent system contains a run for every possible execution of
the protocol, for instance, for every possible key used by the
principals, subject to the restrictions above (such as MP1--MP3).

Formally,
a protocol for agent $i$ is a function
from her local state to the set of actions that she can perform at that
state.  For ease of exposition, the only actions we consider here are
those of sending messages 
(although we could easily
incorporate other actions, such as choosing keys, or tossing coins to
randomize protocols). 
A {\em joint protocol\/}
$P=(P_e,P_1,\ldots,P_n)$, consisting of a protocol for each of the
agents (including a protocol for the environment), associates with
each global state a set of possible {\em joint actions\/} (i.e., tuples
of actions) in the obvious way. Joint actions 
transform global states. To capture their effect, 
we associate with every joint action $\mathsf{a}$ a function
$\tau(\mathsf{a})$ from global states to global states. This function
captures, for instance, the fact that a message sent by an agent will be
received by another agent, and so on. Given a context consisting
of a set of initial global states, an interpretation $\tau$ for
the joint actions, and a protocol $P_e$ for the environment, we can generate
a system corresponding to the joint protocol $P$ in a straightforward
way. Intuitively, the system consists of all the runs $r$ that could
have been 
generated by the joint protocol $P$, that is, for all $m$, $r(m+1)$
is the result of applying $\tau({\mathsf{a}})$ to $r(m)$, where
$\mathsf{a}$ is a joint action that could have been performed
according to the joint protocol $P$ to $r(m)$.%
\footnote{
It is also possible to represent a protocol 
in 
other ways, such as 
in terms of {\em strand spaces\/} \cite{r:thayer99}.  Whichever
representation is used, it should be possible to get a system
corresponding to the protocol.  For example, Halpern and Pucella
\citeyear{r:halpern03d} show how to get a system from a strand space
representation.  
For the purposes of this paper, the precise mechanism used to derive
the multiagent system is not central, although it is an important
issue for the development of formal tools for analyzing protocols.
}

This way of generating systems from protocols is quite general. For 
instance, multiple protocol sessions can be modeled by using an action 
specifically for starting a new session of the protocol, in
combination with protocol action being tagged by the session to which
the action applies.
More generally, most existing approaches to describing systems can be
modeled using these kinds of protocols---it is a simple matter, for
example, to take a process calculus expression and derive a
protocol in our framework that generates the same system as the
original process calculus expression.
(Of course, process calculi come equipped with reasoning techniques
that cannot be easily captured in our framework, but we are concerned
with the ability to represent systems here.)

\section{A Logic for Security Properties}\label{s:logic}

The aim is to be able to reason about properties of security systems
as defined in the last section, 
including properties involving the knowledge of agents in the
system. To formalize this type of reasoning, we first need a
language. 
The logic of algorithmic knowledge \cite[Chapters 10 and 11]{r:fagin95}
provides such a framework.  It extends the classical logic of knowledge 
by adding algorithmic knowledge operators.

The syntax of the logic $\LKXn$ for algorithmic knowledge is
straightforward.  
Starting with a set $\Phi_0$ of
primitive propositions, 
which we can think of as describing basic facts about the system, such
as ``the key is $\ky$'' or ``agent $A$ sent the message $\msg$ to $B$'', 
formulas 
of $\LKXn(\Phi_0)$
are formed by closing off under negation,
conjunction, and the modal operators $K_1$, $\ldots$, $K_n$
and
$X_1,\ldots,X_n$.  
The formula $K_i\phi$ is read as ``agent $i$ (implicitly) knows the
fact $\phi$'', while $X_i\phi$ is read as ``agent $i$ explicitly
knows fact $\phi$''. In fact, we will read $X_i\phi$ as ``agent $i$
can compute fact $\phi$''. This reading will be made precise when we
discuss the semantics of the logic. 
As usual, we take $\phi\vee\psi$ to be an abbreviation for
$\neg(\neg\phi\wedge\neg\psi)$ and $\phi\rimp\psi$ to be an
abbreviation for $\neg\phi\vee\psi$.

The standard models for this logic 
are based on the idea of possible worlds and Kripke structures
\cite{r:kripke63}. Formally, a Kripke structure $M$ is a tuple $(S, \pi, \K_1,
\ldots, \K_n)$, where $S$ is a set of states or possible 
worlds, $\pi$ is an interpretation which associates with each state in $S$
a truth assignment to the primitive propositions (i.e., $\pi(s)(p) \in 
\{\btrue, \bfalse\}$ for each state $s\in S$ and each primitive proposition
$p$), 
and $\K_i$ is an equivalence relation on $S$ (recall that an equivalence
relation is a binary relation which is reflexive, symmetric, and
transitive). $\K_i$ is agent $i$'s possibility relation. Intuitively,
$(s, t) \in \K_i$ if agent $i$ cannot distinguish state $s$ from state $t$ (so that if $s$
is the actual state of the world, agent $i$ would consider $t$ a possible
state of the world). 

A system can be viewed as a Kripke structure, once we add a function
$\pi$ telling us how to assign truth values to the primitive 
propositions. An \emph{interpreted system} $\I$ consists of a pair $(\R,
\pi)$, where $\R$ is a system and $\pi$ is an interpretation for the
propositions in $\Phi$ that assigns truth values to the primitive
propositions at the global states. Thus, for every $p\in\Phi$ and global
state $s$ that arises in $\R$, we have $\pi(s)(p)\in \{\btrue, \bfalse\}$. Of course,
$\pi$ also induces an interpretation over the points of $\R$; simply take
$\pi(r, m)$ to be $\pi(r(m))$. We refer to the points of the system $\R$
as points of the interpreted system $\I$.

The interpreted system $\I=(\R,\pi)$ can be made into a Kripke
structure by taking the possible worlds to be the points of $\R$, and
by defining $\K_i$ so that $((r, m), (r', m'))\in\K_i$ if $r_i(m) =
r'_i(m')$. Clearly $\K_i$ is an equivalence relation on
points. Intuitively, agent $i$ considers a point $(r', m')$ possible
at a point $(r, m)$ if $i$ has the same local state at both
points. Thus, the agents' knowledge is completely determined by their
local states. 

To account for $X_i$, we
provide each agent with a
knowledge algorithm that he uses to compute his knowledge. We will
refer to $X_i\phi$ as \emph{algorithmic knowledge}.  An
\emph{interpreted algorithmic 
knowledge
system} has the form
$(\R,\pi,\A_1,\ldots,\A_n)$, where $(\R,\pi)$ is an interpreted
system and $\A_i$ is the knowledge algorithm of agent $i$. In local
state $\sL$, the agent computes whether he knows $\phi$ by applying
the knowledge algorithm $\A$ to input $(\phi,\sL)$. The output is
either ``Yes'', in 
which case the agent knows $\phi$ to be true, ``No'', in which case
the agent does \emph{not} know $\phi$ to be true, or ``?'', which
intuitively says that the algorithm has insufficient resources to
compute the answer. It is the last clause that allows us to deal with
resource-bounded reasoners. 

We define what it means for a formula $\phi$ to be true (or
satisfied) at a point $(r, m)$ in an interpreted system $\I$, written
$(\I, r, m)\sat\phi$, inductively as follows:
\begin{itemize}
\item[] $(\I, r, m)\sat p$ if $\pi(r,m)(p) = \btrue$
\item[] $(\I,r,m)\sat\neg\phi$ if $(\I,r,m)\not\sat\phi$
\item[] $(\I,r,m)\sat\phi\wedge\psi$ if $(\I,r,m)\sat\phi$ and
$(\I,r,m)\sat\psi$
\item[] $(\I,r,m)\sat K_i\phi$ if $(\I,r',m')\sat\phi$ for all $(r',m')$
such that $r_i(m)=r'_i(m')$
\item[] $(\I,r,m)\sat X_i\phi$ if $\A_i(\phi,r_i(m))=\YES$.
\end{itemize}
The first clause shows how we use the $\pi$ to define the semantics of the
primitive propositions. The next two clauses, which define the
semantics of $\neg$ and $\wedge$, are the standard clauses from propositional
logic. The fourth clause is designed to capture the intuition that
agent $i$ knows $\phi$ exactly if $\phi$ is true in all the worlds
that $i$ thinks are possible. 
The last clause captures the fact that explicit knowledge is
determined using the knowledge algorithm of the agent. 

As we pointed out, we think of $K_i$ as representing \emph{implicit
knowledge}, facts 
that the agent implicitly knows, given its information, while $X_i$
represents \emph{explicit knowledge}, facts whose truth the agent can
compute explicitly.
As is well known, implicit knowledge suffers from the
logical omniscience problem; agents implicitly know all valid formulas
and agents implicitly know all the logical consequences of their
knowledge (that is, $(K_\phi \land K_i (\phi \rimp \psi)) \rimp K_i
\psi$ is valid).  Explicit knowledge does not have that problem.  Note that,
as defined, there is no necessary connection between $X_i\phi$ and
$K_i\phi$. An algorithm could very well claim that agent $i$ knows $\phi$ (i.e., output
``Yes'') whenever it chooses to, including at points where $K_i\phi$ does not
hold. Although algorithms that make mistakes are common, we are often
interested in knowledge algorithms that are correct. 
A knowledge
algorithm is \emph{sound} for agent $i$ in the system $\I$ if for all
points $(r,m)$ of $\I$ and formulas $\phi$, 
$\A(\phi,r_i(m)) = \mbox{``Yes''}$ implies $(\I, r, m)\sat K_i\phi$,
and $\A(\phi,r_i(m)) = \mbox{``No''}$ implies $(\I, r, m) \sat \neg
K_i\phi$. Thus, a knowledge algorithm is sound if its answers are always 
correct.%
\footnote{Note that for the purpose of this paper, there is no need for us
to distinguish between ``No'' and ``?''---none of our algorithms ever
return ``No'. We do keep the distinction, however, as well as the
definition of soundness in cases the algorithm returns ``No'', for
consistency with existing uses of algorithmic knowledge.} 

To reason about security protocols, 
we use the following set $\PhiA$ of primitive propositions:
\begin{iteMize}{$\bullet$}
\item $\sent_i(j,\msg)$: agent $i$ sent message $\msg$ to agent $j$;
\item $\recv_i(\msg)$: agent $i$ received message $\msg$;
\item $\sees_i(\msg)$: agent $i$ has message $\msg$.
\end{iteMize}
Intuitively, $\sent_i(j,\msg)$ is true when agent $i$ has sent message
$\msg$ at some point, intended for agent $j$, and $\recv_i(\msg)$ is
true when agent $i$ has 
received  
message $\msg$ at some point. Agent $i$ \emph{has} a submessage $\msg_1$
at a point $(r,m)$, written $\sees_i(\msg_1)$, if there exists a message
$\msg_2\in\cM$ such that $\recv(\msg_2)$ is in $r_i(m)$, the local state of
agent $i$, and $\msg_1\sub \msg_2$. Note that the $\sees_i$ predicate
is not constrained by encryption.  If $\sees_i(\encr{\msg}{\ky})$
holds, then so does $\sees_i(\msg)$, whether or not agent~$i$ knows the
decryption
key $\ky^{-1}$.  
Intuitively, the $\sees_i$ predicate is true of messages that the
agent considers (possible) part of the messages she has received, as
captured by the $\sub$ relation.

An \emph{interpreted algorithmic knowledge security system} is simply
an interpreted algorithmic knowledge system
$\I=(\R,\pi,\A_1,\dots,\A_n)$, where $\R$ is a security system, 
the set $\Phi_0$ of primitive propositions includes $\PhiA$,
and
$\pi$ is an \emph{acceptable} interpretation, that is, it gives the following
fixed interpretation to the primitive propositions in $\PhiA$: 
\begin{iteMize}{$\bullet$}
\item $\pi(r,m)(\sent_i(j,\msg)) = \btrue$ if and only if
  $\sent(j,\msg)\in r_i(m)$ 
\item $\pi(r,m)(\recv_i(\msg)) = \btrue$ if and only if
$\recv(\msg)\in r_i(m)$ 
\item $\pi(r,m)(\sees_i(\msg)) = \btrue$ if and only if 
there exists $\msg'$ such that $\msg\sub \msg'$ and $\recv(\msg')\in 
r_i(m)$. 
\end{iteMize}
This language can easily express the type of confidentiality (or
secrecy) properties that we focus on here.
Intuitively, we want to guarantee that
throughout a protocol interaction, the adversary does not know a
particular message. Confidentiality properties are stated naturally in
terms of knowledge, for example, ``agent $1$ knows that the key $k$ is
a key known only to agent $2$ and herself''.  Confidentiality
properties are well studied, and central to most of the approaches to
reasoning about security protocols.\footnote{A general
definition of secrecy in terms of knowledge is presented by Halpern
and O'Neill \citeyear{r:halpern02a} in the context of
information flow, a setting that does not take into account
cryptography.} Higher-level security properties, such as
authentication properties, can often be established via
confidentiality properties. See \cite{r:syverson01} for more
details.

To illustrate some of the issues involved, consider an authentication
protocol such as the Needham-Schroeder-Lowe protocol
\cite{r:lowe95}. A simplified version of the protocol is characterized
by the following message exchange between two agents $A$ and $B$:
\[\begin{array}{l}
A \rightarrow B: \encr{n_A,A}{k_B}\\
B \rightarrow A: \encr{n_A,n_B,B}{k_A}\\
A \rightarrow B: \encr{n_B}{k_B}.
\end{array}\]
An authentication property of this protocol can be expressed
informally as follows: under suitable assumptions on the keys known to
the adversary and the fact that $B$ is running his part of the
protocol, $A$ knows that $n_A$ and $n_B$ are kept confidential between
her and $B$.\footnote{It may be more reasonable to
talk about {\em belief\/} rather than {\em knowledge\/} that
$n_A$ and $n_B$ are kept confidential.  For simplicity, we talk about
knowledge in this paper.  Since most representations of belief suffer
from logical omniscience, what we say applies to belief as well as
knowledge.} 
From this, she knows that she is interacting with $B$,
because she has received a message containing $n_A$, which only $B$
could have produced. Similarly, $A$ also knows that when $B$ receives
her message, $B$ will know that he is interacting with $A$, because
only $A$ knows the nonce $n_B$ which is part of the last message.
Similar reasoning can be applied to $B$. This argument relies on the
confidentiality of the nonces $n_a$ and $n_b$. 
It is tempting to capture this fact by stating that no agent $i$
other than $A$ and $B$ knows $\sees_i(n_A)$ or $\sees_i(n_B)$.
(We must write this as $K_i\sees_i(n_A)$ or $K_i\sees_i(n_B)$ because
we cannot express directly the idea of knowing a message---an agent
can only know if they have received a message, or if a message is a
possible component of a message they have received.)

Unfortunately, because the implicit knowledge operator suffers
from logical omniscience, such a statement does not capture the intent
of confidentiality. 
At every point where an adversary $a$ intercepts a message
$\encr{n_A,n_B,B}{k_A}$, $K_a\sees_a(n_A)$ is true (since
$n_A\sub\encr{n_A,n_B,B}{k_A}$), and hence the adversary knows
that he has seen the nonce $n_A$, irrespective of whether 
he knows the decryption key corresponding to $k_A$). 
This shows that the 
standard interpretation of knowledge expressed via the 
implicit knowledge operator does not capture
important aspects of reasoning about security.
The adversary having the implicit knowledge
that $n_A$ is part of the message does not suffice, in general, for the
adversary to
\emph{explicitly} know that $n_A$ is part of the message. 
Intuitively, the adversary may not have the capabilities to realize he
has seen $n_A$.

A more reasonable interpretation of confidentiality in this 
setting is $\neg X_a\sees_a(n_A)$: the adversary does not
explicitly know (i.e., cannot compute or derive) whether he has seen the
nonce $n_A$. Most logics of security, instead of relying on a notion
of knowledge, introduce special primitives to
capture the fact that the adversary can see a message $m$ encrypted
with key $k$ only if he has access to the key $k$.  Doing this
hardwires the capabilities of the adversary into the
semantics of the logic. Changing these capabilities requires changing the
semantics. In our case, we simply need to supply the appropriate
knowledge algorithm to the adversary, capturing his capabilities. In
the following section, we examine in more detail the kind of knowledge 
algorithms that correspond to interesting capabilities.

\section{Modeling Adversaries}\label{s:adversaries}
As we showed in the last two sections, interpreted algorithmic
knowledge security systems can be used to provide a foundation for
representing 
security protocols, and support a logic for writing properties based
on knowledge, both 
traditional (implicit) and algorithmic (explicit).  
For the purposes of analyzing security protocols, we use traditional
knowledge to model a principal's beliefs about what can happen in the
protocol, while we use algorithmic knowledge to model the adversary's
capabilities, possibly resource-bounded. 
To interpret algorithmic knowledge, we rely on a
knowledge algorithm for each agent in the system. 
We use the adversary's knowledge algorithm to capture the adversary's
ability to draw conclusions from what he has seen.
In this section, we show
how we can capture different capabilities for the adversary 
in a natural way
in this framework.  We first show how to capture the
standard model of adversary due to Dolev and Yao. We then show how to
account for the adversary in the Duck-Duck-Goose protocol, 
and
the adversary considered by Lowe \citeyear{r:lowe02} that can perform
self-validating guesses. 

We start by considering passive (or eavesdropping) adversaries,
which simply record every message exchanged by the principals;
in Section~\ref{s:active}, we consider active adversaries.
For
simplicity, we assume a single adversary per system; our results
extend to the general case immediately, but the notation 
becomes cumbersome. 

Adversaries generally have the ability to eavesdrop on all
communications, and in the case of active adversaries, to furthermore
intercept messages and forward them at will. 
To model both the eavesdropping of a message and its interception, we
assume an internal action $\intercept(\msg)$ meant to capture the fact
that an adversary has intercepted message $\msg$. 
Because messages may or may not be delivered to their final
destination, this models both eavesdropping---in which case the message
does in fact get received by its intended recipient---and actual
interception---in which case the message does not get received by its
intended recipient. 
Because an intercepted message is available to an agent in its local
state, we adapt the notion of an acceptable interpretation from
Section~\ref{s:logic} to allow intercept messages to be used for
determining which messages an agent has:
\begin{iteMize}{$\bullet$}
\item $\pi(r,m)(\sees_i(\msg)) = \btrue$ if and only if 
there exists $\msg'$ such that $\msg\sub \msg'$ and either $\recv(\msg')\in 
r_i(m)$ or $\intercept(\msg')\in r_i(m)$.
\end{iteMize}

\subsection{Passive Adversaries}\label{s:passive}
Passive adversaries can be modeled formally as
follows. An \emph{interpreted algorithmic knowledge security system with
passive adversary $a$ ($a\in\{1,\dots,n\}$)}  is
an interpreted algorithmic knowledge security system
$\I=(\R,\pi,\A_1,\dots,\A_n)$ 
satisfying the following constraints at all points $(r,m)$:
\begin{axiomlist}
\item[P1.] $r_a(m)$ consists only of $\intercept(\msg)$ events.
\item[P2.] 
For all $j$ and events $\sent(j,\msg)$ in $r_j(m)$, there exists an event
$\intercept(\msg)$ in $r_a(m)$.
\item[P3.] For every $\intercept(\msg)$ in $r_a(m)$, there is a
  corresponding $\sent(j,\msg)$ in $r_i(m)$ for some $i$.
\item[P4.] For all $i$ and 
every $\sent(j,\msg)$ in $r_i(m)$, $j\ne a$.
\end{axiomlist}
P1 captures the passivity of the adversary---he can only intercept
messages, not send any; P2 says that every message sent by a principal
is basically copied to the adversary's local state, while P3 says that
only messages sent by principals appear in the adversary's local
state; P4 ensures that a
passive adversary is not an agent to which other agents may
intentionally send messages to (i.e., the adversary is an outsider in
the system).
We next consider various knowledge algorithms for the adversary.

\subsubsection{The Dolev-Yao Adversary}\label{s:dy}

Consider the standard
Dolev-Yao adversary \cite{r:dolev83}. This model is a combination
of assumptions on the encryption scheme used and the capabilities of the
adversaries. Specifically, the encryption scheme is seen as the free algebra
generated by $\cP$ and $\cK$ over operations $\cdot$ and
$\encr{}{}$. Perhaps the easiest way to formalize this is to view the
set $\cM$ as the set of expressions generated by the grammar
\[ \msg ::= \pl ~|~ \ky ~|~ \encr{\msg}{\ky} ~|~ \msg\cdot \msg\]
(with $\pl\in\cP$ and $\ky\in\cK$). 
We assume 
that there are no
collisions; messages always have a unique decomposition.  The only way
that $\encr{\msg}{\ky} = \encr{\msg'}{\ky'}$ is if $\msg = \msg'$ and
$\ky = \ky'$. 
We 
also
make the standard assumption that concatenation and encryption have
enough redundancy to recognize that a term is in fact a concatenation
$\msg_1\cdot\msg_2$ or an encryption $\encr{\msg}{\ky}$.

The 
classical
Dolev-Yao model can be formalized by a relation $H\derivesDY \msg$
between a set $H$ of messages and a message $\msg$.
(Our formalization is equivalent to many other formalizations of Dolev-Yao
in the literature, and is similar in spirit to that of Paulson
\citeyear{r:paulson98}.) 
Intuitively, $H\derivesDY \msg$ means that an adversary can
``extract'' message $\msg$ 
from a set of received messages and keys $H$, using the allowable
operations. The derivation is defined using the following inference
rules:
\[ \Rule{\msg\in H}{H\derivesDY \msg} \quad \Rule{H\derivesDY\encr{\msg}{\ky}
\quad H\derivesDY \ky^{-1}}{H\derivesDY \msg} \quad
\Rule{H\derivesDY \msg_1\cdot \msg_2}{H\derivesDY \msg_1} \quad
\Rule{H\derivesDY \msg_1\cdot \msg_2}{H\derivesDY \msg_2}.\]
This presentation of the Dolev-Yao capabilities is restricted, because
it only allows an adversary to deconstruct messages, and not to
construct them. 
In many situations with passive adversaries---that is, adversaries that
cannot inject new messages into the system---this is not a significant
restriction. 
It is easy to extend the relation to allow for the construction of new
messages using concatenation and encrytion. 
(Of course, extending the relation in such a way would require a
corresponding modification to the knowledge algorithm below.)

\newcommand{\submsg}{\mathit{partof}}

\begin{figure*}[t]
\hrule
\begin{align*}
\submsg (\msg,\msg',K):\ \ &
\begin{prog}
 \mbox{if $\msg=\msg'$ then}\\
 \tspace \mbox{return $\mathit{true}$}\\
 \mbox{if $\msg'$ is $\encr{\msg_1}{\ky}$ and $\ky^{-1}\in K$ then}\\
 \tspace \mbox{return $\submsg(\msg,\msg_1,K)$}\\
 \mbox{if $\msg'$ is $\msg_1\cdot \msg_2$ then}\\
 \tspace \mbox{return $\submsg(\msg,\msg_1,K) \lor \mathit{submsg(\msg,\msg_2,K)}$}\\
 \mbox{return $\mathit{false}$}
\end{prog}\\[.2in]
\mathit{getkeys} (\msg,K):\ \ & 
 \begin{prog}
  \mbox{if $\msg\in\cK$ then}\\
 \tspace \mbox{return $\{\msg\}$}\\
 \mbox{if $\msg'$ is $\encr{\msg_1}{\ky}$ and $\ky^{-1}\in K$ then}\\
 \tspace \mbox{return $\mathit{getkeys} (\msg_1,K)$}\\
 \mbox{if $\msg'$ is $\msg_1\cdot \msg_2$ then}\\
 \tspace \mbox{return $\mathit{getkeys}(\msg_1,K)\cup\mathit{getkeys}(\msg_2,K)$}\\
 \mbox{return $\{\}$}
 \end{prog}\\[.2in]
\mathit{keysof} (\sL):\ \ & 
 \begin{prog}
  \mbox{$K \leftarrow \mathit{initkeys}(\sL)$}\\
 \mbox{loop until no change in $K$}\\
 \tspace \mbox{$K \leftarrow \bigcup\limits_{\recv(\msg)\in\sL}\mathit{getkeys}(\msg,K)$}\\
 \mbox{return $K$}
 \end{prog}
\end{align*}
\caption{Dolev-Yao knowledge algorithm auxiliary functions}
\label{f:ady}
\medskip\hrule
\end{figure*}

In our framework, to capture the capabilities of a Dolev-Yao
adversary,
we specify how the adversary can explicitly know that she \emph{has} a
message, by 
defining a knowledge algorithm $\ADYi$ for adversary $i$.  Recall that
a knowledge algorithm for agent $i$ takes as input a formula and agent
$i$'s local state (which we are assuming contains the messages
received by $i$).  The most interesting case in the definition of
$\ADYi$ is when the formula is $\sees_i(\msg)$.  To compute
$\ADYi(\sees_i(\msg),\ell)$, the algorithm simply checks, for every
message $\msg'$ received by the adversary, whether $\msg$ is a submessage of
$\msg'$, according to the keys that are known to the adversary.  
We assume that the adversary's initial state consists of
the set of keys initially known by the adversary.
This will typically contain, in a public-key
cryptography setting, the public keys of all the agents.  
We use $\mathit{initkeys}(\ell)$ to denote the set of initial keys known
by agent $i$ in local state $\ell$. (Recall that a local state
for agent $i$ is the sequence of events pertaining to agent $i$,
including any initial information in the run, in this case, the keys
initially known.) 
The function $\submsg$, which can take apart messages created by
concatenation, or decrypt messages as long as the adversary knows
the decryption key, is used to check whether $\msg$ is a submessage of
$\msg'$.
$\ADYi(\sees_i(\msg),\sL)$ is defined as follows:
\[ 
 \begin{prog}
 \mbox{if $\msg\in\mathit{initkeys}(\ell)$ then return ``Yes''}\\
 \mbox{$K \leftarrow \mathit{keysof}(\sL)$}\\
 \mbox{for each $\recv(\msg')$ and $\intercept(\msg')$ in $\sL$}\\
 \tspace \mbox{if $\submsg(\msg,\msg',K)$ then}\\
 \tspace \tspace \mbox{return ``Yes''}\\
 \mbox{return ``?''.}
 \end{prog}\]
The auxiliary functions used by the algorithm are given in
Figure~\ref{f:ady}.  
In particular, the function $\submsg$ captures the $\sub$ relation.

In the Dolev-Yao model, an adversary cannot explicitly
compute anything interesting about what other messages agents have.
Hence, for other primitives, including $\sees_j(\msg)$ for $j\ne i$,
$\ADYi$ returns \DONTK.  For formulas of the form $K_j \phi$ and $X_j
\phi$, $\ADYi$ also returns \DONTK.  For Boolean combinations of
formulas, $\ADYi$
returns the corresponding Boolean combination (where
the negation of $\DONTK$ is $\DONTK$, the conjunction of $\NO$ and
$\DONTK$ is $\NO$, and the conjunction of $\YES$ and $\DONTK$ is
$\DONTK$) of the answer for each $\sees_i(\msg)$ query.

The following result shows that an adversary using $\ADYi$ recognizes
(i.e., returns $\YES$ to) $\sees_i(\msg)$ in state $\ell$ if and only if $\msg$
is one of the messages that can
be derived (according to $\derivesDY$) from the messages received in
that state
together with the keys initially known,
Moreover, if a $\sees_i(\msg)$ formula is derived at the point $(r,m)$,
then $\sees_i(\msg)$ is actually true at $(r,m)$ (so that $\ADYi$ is sound).
\begin{proposition}\label{p:dolevyao}
Let $\I = (\R,\pi,\A_1,\ldots,\A_n)$ be an interpreted algorithmic
knowledge security system where $\A_i = \ADYi$. Then 
\begin{multline*}
\text{$(\I,r,m) \sat X_i(\sees_i(\msg))$ if and only if}\\
\{ \msg' : \recv(\msg')\in r_i(m)\} \union\{\msg' :
\intercept(\msg')\in r_i(m)\} 
\union \mathit{initkeys}(\ell) \derivesDY \msg.
\end{multline*}
Moreover, if  
$(\I,r,m) \sat X_i(\sees_i(\msg))$ then $(\I,r,m) \sat \sees_i(\msg)$.
\end{proposition}
\begin{proof}
Let $K=\mathit{keysof}(r_i(m))$.
We must show that $\ADY_i(\sees_i(\msg), r_i(m)) = \YES$ if and only if 
$K \union \{ \msg' : \recv(\msg')\in r_i(m)\}\union\{\msg' :
  \intercept(\msg')\in r_i(m)\} \derivesDY \msg$.
It is immediate from the description of $\ADY_i$ and $\derivesDY$ that
this is true if $\msg\in\mathit{initkeys}(r_i(m))$.  If
$\msg\notin\mathit{initkeys}(r_i(m))$, 
then $\ADY_i(\sees_i(\msg), r_i(m)) = \YES$ if and only if 
$\submsg (\msg,\msg',K) = \mathit{true}$ for some $\msg'$ such that 
$\recv(\msg')\in r_i(m)$ or $\intercept(\msg')\in r_i(m)$.  Next
observe that $\submsg (\msg,\msg',K) = \mathit{true}$ if and only if
$K\cup\{\msg'\} \derivesDY \msg$: the 
``if'' direction follows by a simple induction on the length of the
derivation; the ``only if'' direction follows by a straightforward
induction on the structure of $\msg$.  Finally, observe that if $M$ is a
set of messages, then $K \union M \derivesDY \msg$ if and only if 
$K \union \{\msg'\} \derivesDY \msg$ for some $\msg' \in M$.   The
``if'' direction is trivial.  The ``only if'' direction follows by
induction on the number of times the rule ``from $\msg' \in H$ infer 
$H \derivesDY \msg'$'' is used to derive some $\msg' \in M$.  
If it is never used, then
it is easy to see that $K \derivesDY \msg'$.  If it is used more than
once, and the last occurrence is used to derive $\msg'$, then it is easy
to see that $K \union \{\msg'\} \derivesDY \msg'$ (the derivation just
starts from the last use of this rule).  The desired result is now
immediate. 
\end{proof}

In particular, if we have an interpreted algorithmic knowledge
security system with a passive adversary $a$ such that $\A_a=\ADY_a$,
then Proposition~\ref{p:dolevyao} captures the knowledge of a passive
Dolev-Yao adversary.

\subsubsection{The Duck-Duck-Goose Adversary}

The key advantage of our framework is that we can easily change the
capabilities of the adversary beyond those prescribed by the Dolev-Yao
model.  For example, we can capture the fact that if the adversary
knows the protocol, she can derive more information than she could
otherwise. For instance, in the Duck-Duck-Goose example, 
assume that the adversary maintains in her local state a list of all
the bits received corresponding to the key of the principal.
We can 
write the algorithm so that if the adversary's local state contains all
the bits of the key of the principal, then the adversary can decode
messages that have been encrypted with that key.
Specifically, assume that key $\ky$ is being sent in the Duck-Duck-Goose
example. Then for an adversary $i$, $\sees_i(\ky)$ will be false until all
the bits of 
the key have been received. This translates immediately into the
following algorithm $\ADDGi$:
\[ 
  \begin{prog}
  \text{if all the bits recorded in $\sL$ form $\ky$ then}\\
  \tspace \text{return ``Yes'' else return ``?''.}
  \end{prog} \]
$\ADDGi$ handles other formulas in the same way as $\ADYi$.

Of course, nothing keeps us from combining algorithms, so that we can
imagine an adversary intercepting both messages and key bits, and using 
an algorithm $\A_i$ that is a combination of the Dolev-Yao algorithm
and the Duck-Duck-Goose algorithm; $\A_i(\phi,\sL)$ is defined as follows:
$$\begin{array}{l}
\mbox{if $\ADYi(\phi,\sL)=\mbox{``Yes''}$ then}\\
     \tspace \mbox{return ``Yes''}\\
     \mbox{else return $\ADDGi(\phi,\sL)$.}
\end{array}$$
This assumes that the adversary knows the protocol, and hence knows
when the key bits are being sent. The algorithm above captures
this protocol-specific knowledge.

To see why this adversary is not completely trivial, note that the
obvious way of trying to capture this kind of adversary in a Dolev-Yao
model, where we allow operations for combining keys to form new keys,
fails in that the resulting adversary is too powerful. 
If we assume that keys are formed out of bits, and we allow a
bit-concatenation operation that lets the adversary create keys out
of bits, then a Dolev-Yao adversary with such extended operations, as
soon as they have received bits 1 and 0, would be able to construct any
possible key, and therefore such an adversary would be able to decrypt
any encrypted message.

\subsubsection{The Lowe Adversary}\label{s:lowemodel}

For a more realistic example 
of an adversary model that goes beyond
Dolev-Yao, consider the following adversary model introduced by Lowe
\citeyear{r:lowe02} to analyze protocols subject to guessing
attacks. The intuition is that some protocols provide for a way to
``validate'' the guesses of an adversary. For a simple example of this, 
here is a simple challenge-based authentication protocol:
\[\begin{array}{l}
A \rightarrow S: A\\
S \rightarrow A: n_s\\
A \rightarrow S: \encr{n_s}{p_a}.
\end{array}\]
Intuitively, $A$ tells the server $S$ that she wants to authenticate
herself. $S$ replies with a challenge $n_s$. $A$ sends back to $S$ the
challenge encrypted with her password $p_a$. Presumably, $S$ knows the
password, and can verify that she gets
$\encr{n_s}{p_a}$. Unfortunately, an adversary can overhear both $n_s$
and $\encr{n_s}{p_a}$, and can ``guess'' a value $g$ for $p_a$ and
verify his guess by checking if $\encr{n_s}{g}=\encr{n_s}{p_a}$.  The
key feature of this kind of attack is that the guessing (and the
validation) can be performed offline, based only on the intercepted
messages. 

\newcommand{\lowesingle}{\rhd}
\newcommand{\lowedouble}{\unrhd}

To account for this capability of adversaries is actually fairly
complicated. 
We present a slight variation of Lowe's description, mostly to
make it notationally consistent with the rest 
of the section; we refer the reader to Lowe \citeyear{r:lowe02} for a
discussion of the design choices. 

Lowe's model relies on a basic one-step reduction function, 
$S\lowesingle_l \msg$, saying that the messages in $S$ can be used to
derive the message $\msg$. 
Its definition is reminiscent of 
$\derivesDY$, except that it represents a single step of
derivation. 
Note that the derivation relation $\lowesingle_l$ is ``tagged'' by the
kind of derivation performed ($l$).
\begin{align*}
\{\msg,\ky\} & \lowesingle_\mathsf{enc}\encr{\msg}{\ky}\\
\{\encr{\msg}{\ky},\ky^{-1}\} & \lowesingle_\mathsf{dec} \msg\\
\{\msg_1\cdot \msg_2\} & \lowesingle_\mathsf{fst}\msg_1\\
\{\msg_1\cdot \msg_2\} & \lowesingle_\mathsf{snd}\msg_2.
\end{align*}
Lowe also includes a reduction to derive $\msg_1\cdot \msg_2$ from $\msg_1$
and $\msg_2$. We do not add this reduction to simplify the
presentation. 
It is straightforward to extend our approach to deal with it.

Given a set $H$ of messages, and a sequence $t$ of one-step reductions, 
we define inductively the set $[H]_t$ of messages obtained from the
one-step reductions given in $t$:
\begin{align*}
[H]_{\langle\rangle} & =  H\\
[H]_{\langle S\lowesingle_l \msg\rangle\cdot t} & = 
  \begin{cases}
  [H\cup\{\msg\}]_t & \text{if $S\subseteq H$}\\
  \mathit{undefined} & \text{otherwise.}
  \end{cases}
\end{align*}
Here, $\<\>$ denotes the empty trace, and $t_1\cdot t_2$ denotes
trace concatenation.
A trace $t$ is said to be \emph{monotone} if, intuitively, it does not 
perform any one-step reduction that ``undoes'' a previous
one-step reduction. For example, the reduction
$\{\msg,\ky\}\lowesingle\encr{\msg}{\ky}$ undoes the reduction
$\{\encr{\msg}{\ky},\ky^{-1}\}\lowesingle \msg$. (See Lowe
\citeyear{r:lowe02} for more details on undoing reductions.) 

We say that a set $H$ of messages \emph{validates} a guess $\msg$ 
if $H$ contains enough information to verify that $\msg$
is indeed a good guess. Intuitively, this happens if a value $v$
(called a validator) can be derived from the messages in
$H\cup\{\msg\}$ in a way that uses the guess $\msg$, and either that
(a) validator $v$ can be derived
in a different way from $H\cup\{\msg\}$, (b) the validator $v$ is already
in $H\cup\{\msg\}$, or (c) the validator $v$ is a key whose inverse is
derivable from $H\cup\{\msg\}$. For example, in the protocol exchange
at the beginning of this section, the adversary sees the messages
$H=\{n_s,\encr{n_s}{p_a}\}$, and we can check that $H$ validates the
guess $\msg=p_a$:~clearly,
$\{n_s,\msg\}\lowesingle_\mathsf{enc}\encr{n_s}{p_a}$, and 
$\encr{n_s}{p_a}\in H\cup\{\msg\}$. In this case, the validator
$\encr{n_s}{p_a}$ is already present in $H\cup\{\msg\}$. For other
examples of validation, we again refer to Lowe \citeyear{r:lowe02}.

We can now define the relation $H\derivesL \msg$ that says that $\msg$ can be 
derived from $H$ by a Lowe adversary. Intuitively, $H\derivesL \msg$ if
$\msg$ can be derived by Dolev-Yao reductions, or $\msg$ can be guessed and
validated by the adversary, and hence susceptible to an
attack. 
Formally, $H\derivesL \msg$ if and only if $H\derivesDY \msg$ or there
exists a monotone trace $t$, a set $S$, and a ``validator'' $v$ such that
\begin{enumerate}[(1)]
\item $[H\cup\{\msg\}]_t$ is defined;
\item $S\lowesingle_l v$ is in $t$;
\item there is no trace $t'$ such that $S\subseteq[H]_{t'}$; and
\item either:
\begin{enumerate}[(a)]
\item there exists $(S',l')\ne(S,l)$ with $S'\lowesingle_{l'} v$
in $t$
\item $v\in H\cup\{\msg\}$ or 
\item $v\in\cK$ and $v^{-1}\in [H\cup\{\msg\}]_t$. 
\end{enumerate}
\end{enumerate}
It is not hard to verify that this formalization captures the intuition
about validation given earlier. Specifically, condition (1) says that
the trace $t$ is well-formed, condition (2) says that the validator
$v$ is derived from $H\cup\{\msg\}$, condition (3) says that deriving the 
validator $v$ depends on the guess $\msg$, and condition (4) specifies
when a validator $v$ validates a guess $\msg$, as given earlier. 

\begin{figure*}[tp]
\hrule
\begin{align*}
\mathit{guess}(\msg,\sL):\ \  & 
 \begin{prog}
 \mbox{$H \leftarrow \mathit{reduce}(\{\msg' : \recv(\msg')\in\sL
   ~\mathit{or}~ \intercept(\msg')\in\sL\} \cup\mathit{initkeys}(\sL))\cup\{\msg\}$}\\
 \mbox{$\mathit{reds} \leftarrow \{\}$}\\
 \mbox{loop until $\mathit{reductions}(H)-\mathit{reds}$ is empty}\\
 \tspace \mbox{$(S,l,v) \leftarrow \mbox{pick an element of } \mathit{reductions} (H)-\mathit{reds}$}\\
 \tspace \mbox{if $\exists(S',l',v)\in\mathit{reds}$ s.t. $S'\ne S$ and $l'\ne l$ then}\\
 \tspace\tspace \mbox{return $\YES$}\\
 \tspace \mbox{if $v\in H$ then}\\
 \tspace\tspace \mbox{return $\YES$}\\
 \tspace \mbox{if $v\in\cK$ and $v^{-1}\in H$ then}\\
 \tspace\tspace \mbox{return $\YES$}\\
 \tspace \mbox{$\mathit{reds} \leftarrow \mathit{reds}\cup\{(S,l,v)\}$}\\
 \tspace \mbox{$H \leftarrow H\cup\{v\}$}\\
 \mbox{return $\NO$}
 \end{prog}\\[1ex]
\mathit{reduce} (H):\ \  & 
  \begin{prog}
   \mbox{loop until no change in $H$}\\
 \tspace \mbox{$r\leftarrow \mathit{reductions}(H)$}\\
 \tspace \mbox{for each $(S,l,v)$ in $r$}\\
 \tspace\tspace \mbox{$H\leftarrow H\cup\{v\}$}\\
 \mbox{return $H$}
  \end{prog}\\[1ex]
\mathit{reductions}(H):\ \  & 
 \begin{prog}
  \mbox{$\mathit{reds}\leftarrow \{\}$}\\
 \mbox{for each $\msg_1\cdot \msg_2$ in $H$}\\
 \tspace \mbox{$\mathit{reds}\leftarrow \{(\{\msg\},\mathsf{fst},\msg_1), (\{\msg\},\mathsf{snd},\msg_2)\}$}\\
 \mbox{for each $\msg_1,\msg_2$ in $H$}\\
 \tspace \mbox{if $\msg_2\in\cK$ and $\mathit{sub}(\encr{\msg_1}{\msg_2},H)$ then}\\
 \tspace\tspace \mbox{$\mathit{reds}\leftarrow \{(\{\msg_1,\msg_2\},\mathsf{enc},\encr{\msg_1}{\msg_2})\}$}\\
 \tspace \mbox{if $\msg_1$ is $\encr{\msg'}{\ky}$ and $\msg_2$ is $\ky^{-1}$ then}\\
 \tspace\tspace \mbox{$\mathit{reds}\leftarrow \{(\{\msg_1,\msg_2\},\mathsf{dec},\msg')\}$}\\
 \mbox{return $\mathit{reds}$}
 \end{prog}\\[1ex]
\mathit{sub}(\msg,H):\ \  & 
  \begin{prog}
  \mbox{if $H=\{\msg\}$ then}\\
 \tspace \mbox{ return $\mathit{true}$}\\
 \mbox{if $H=\{\msg_1\cdot \msg_2\}$ then}\\
 \tspace\tspace \mbox{return $\mathit{sub}(\msg,\{\msg_1\})\vee\mathit{sub}(\msg,\{\msg_2\})$}\\
 \mbox{if $H=\{\encr{\msg'}{\ky}\}$ then}\\
 \tspace\tspace \mbox{return $\mathit{sub}(\msg,\{\msg'\})$}\\
 \mbox{if $|H|>1$ and $H=\{\msg'\}\cup H'$ then}\\
 \tspace\tspace \mbox{return $\mathit{sub}(\msg,\{\msg'\})\vee\mathit{sub}(\msg,H')$}\\
 \mbox{return $\mathit{false}$}
  \end{prog}
\end{align*}
\caption{Lowe knowledge algorithm auxiliary functions}
\label{f:al}
\medskip\hrule
\end{figure*}

We would now like to define a knowledge algorithm $\ALi$ to
capture the capabilities of the Lowe adversary.
Again, the only case of real interest is what $\ALi$ does on input
$\sees_i(\msg)$.
$\ALi(\sees_i(\msg),\sL)$ is defined as follows:
\[ 
  \begin{prog}  
   \mbox{if $\ADYi(\sees_i(\msg),\sL)=\mbox{``Yes''}$ then}\\
 \tspace \mbox{return ``Yes''}\\
 \mbox{if $\mathit{guess}(\msg,\sL)$ then}\\
 \tspace \mbox{return ``Yes''}\\
 \mbox{return ``?''.}
  \end{prog}\]
The auxiliary functions used by the algorithm are given in
Figure~\ref{f:al}.  
(We have not concerned ourselves with matters of efficiency in the
description of $\ALi$; again, see Lowe \citeyear{r:lowe02} for a
discussion of implementation issues.)

As before, we can check the correctness and soundness of the
algorithm: 
\begin{proposition}\label{p:lowe}
Let $\I = (\R,\pi,\A_1,\ldots,\A_n)$ be an interpreted algorithmic knowledge security system  where $\A_i = \ALi$. Then
\begin{multline*}
\text{$(\I,r,m) \sat X_i(\sees_i(\msg))$ if and only if}\\
\{ \msg' : \recv(\msg')\in r_i(m)\} \union \{\msg' :
\intercept(\msg')\in r_i(m)\} \union \mathit{initkeys}(\ell) \derivesL
\msg.
\end{multline*}
Moreover, if 
$(\I,r,m) \sat X_i(\sees_i(\msg))$ then $(\I,r,m) \sat \sees_i(\msg)$.
\end{proposition}
\begin{proof}
Let $K=\mathit{keysof}(r_i(m))$. 
The proof is similar in spirit to that of Proposition~\ref{p:dolevyao},
using the fact that 
if $\msg\not\in\mathit{initkeys}(r_i(m))$ and
$K\cup\{\msg'\mid\recv(\msg')\in r_i(m)\}\cup\{\msg'\mid\intercept(\msg')\in r_i(m)\}\not\derivesDY\msg$, then
$\mathit{guess}(\msg,r_i(m))=\YES$ if and only if
$K\cup\{\msg'\mid\recv(\msg')\in r_i(m)\}\cup\{\msg'\mid\intercept(\msg')\in r_i(m)\}\derivesL \msg$. 
The proof of this fact is essentially given 
by Lowe
\citeyear{r:lowe02}, the algorithm $\ALi$ being a direct translation
of the CSP process implementing the Lowe adversary. 
Again, soundness with respect to $\sees_i(m)$ follows easily.
\end{proof}

\subsection{Active Adversaries}\label{s:active}

Up to now we have considered passive adversaries, which can
intercept messages exchanged by protocol participants, but cannot
actively participate in the protocol.  Passive adversaries are often
appropriate when the concern is confidentiality of messages.  
However, there are many attacks on security protocols that do not
necessarily involve a breach of confidentiality. For instance, some
authentication properties are concerned with ensuring that no
adversary can pass himself off as another principal. This presumes
that the adversary is able to interact with other principals. 
Even when it comes to confidentiality, there are clearly attacks 
that an active adversary can make that cannot be made by a passive
adversary.  

To analyze active adversaries, we need to consider what messages they
can send.  This, in turn depends on 
their capabilities, which we already have captured using
knowledge algorithms. Formally, at a local state $\ell$, an adversary
using knowledge algorithm $\A_i$ can construct the messages 
in the set $C(\ell)$, 
defined to be the closure under $\mathit{conc}$ and $\mathit{encr}$ of 
the set $\{\msg\mid \A_i(\sees_i(\msg),\ell)=\YES\}$ of messages 
that adversary $i$ has. 
For more complex cryptosystems, the construction and destruction of
messages to determine which can in fact be created by the
adversary will be more complex---Paulson~\citeyear{r:paulson98}, for
instance, defines a general approach based on two operations,
analysis and synthesis, interleaved to generate all possible
messages that an adversary can construct. This technique can be
readily adapted to our framework.

Once we consider active adversaries, we must consider whether they are
insiders or outsiders.
Intuitively, an insider is an
adversary that other agents know about, and can initiate interactions
with. (Insider adversaries are sometimes called {\em corrupt principals}
or {\em dishonest principals}.) 
As we mentioned in the introduction, the difference between insiders and
outsiders was highlighted by Lowe's \citeyear{r:lowe95} man-in-the-middle
attack of the Needham-Schroeder protocol.
An \emph{interpreted algorithmic knowledge security system with
active (insider) adversary $a$ ($a\in\{1,\dots,n\}$)}
is an interpreted algorithmic knowledge
security system $\I=(\R,\pi,\A_1,\dots,\A_n)$ 
satisfying the following constraints at all points $(r,m)$.
\begin{axiomlist}
\item[A1.] For every $\intercept(\msg)\in r_a(m)$, there is a corresponding 
$\sent(j,\msg)$ in $r_i(m)$ for some $i,j$.
\item[A2.] For every $\sent(j,\msg)\in r_a(m)$, we have $\msg\in
C(r_a(m))$.
\end{axiomlist}
A1 says that every message sent by the agents can be intercepted by
the adversary and end up in the adversary's local state, and every
intercepted message in the adversary's local state is a message that
has been sent by an agent. 
A2 says that every message sent by the adversary must have been
constructed out of the messages in his local state according to his
capabilities. 
(Note that the adversary can forge the ``send'' field of the
messages.)

To accommodate outsider adversaries, it suffices to add the restriction 
that no message is sent directly to the adversary. Formally, an
\emph{interpreted algorithmic knowledge security system with
active (outsider) adversary $a$ ($a\in\{1,\dots,n\}$)}
is an interpreted algorithmic knowledge
security system $\I=(\R,\pi,\A_1,\dots,\A_n)$ with an active insider
adversary $a$
such that for all points $(r,m)$ and for 
all agents $i$, the following additional constraint is satisfied.
\begin{axiomlist}
\item[A3.] For every $\sent(j,\msg)\in r_i(m)$, $j\ne a$.
\end{axiomlist}

\section{Related Work}\label{s:related}

The issues we raise in this paper are certainly not new, and have been
addressed, up to a point, in the literature. In this section, we
review this literature, and discuss where we stand with respect to
other approaches that have attempted to tackle some of the same
problems. 

As we mentioned in the introduction, the Dolev-Yao adversary is the
most widespread adversary in the literature. Part of its attraction is
its tractability, making it possible to develop formal systems to
automatically check for safety with respect to such adversaries
\cite{r:millen87,r:mitchell97,r:paulson98,r:lowe98,r:meadows96}. The idea of
moving beyond the Dolev-Yao adversary is not new. As we pointed out in 
Section~\ref{s:lowemodel}, Lowe \citeyear{r:lowe02} developed an
adversary that can encode some amount of off-line guessing; we 
showed in Section~\ref{s:lowemodel} that we could capture such an
adversary in our framework. 
More recent techniques for detecting off-line guessing attacks
\cite{r:corin05,r:baudet05} can also be similarly modeled. 
Other approaches have the possibility of extending the adversary model. For
instance, the framework of Paulson \citeyear{r:paulson98}, Clarke,
Jha and Morrero \citeyear{r:clarke98}, and Lowe \citeyear{r:lowe98}
describe the adversary via a 
set of derivation rules, which could be modified by adding new
derivation rules. 
We could certainly capture these adversaries by
appropriately modifying our $\ADYi$ knowledge algorithm.  
(Pucella \citeyear{r:pucella06c} studies the properties
of algorithmic knowledge given by derivation rules in more depth.)
However, these other approaches do not seem to have the
flexibility of our approach in terms of capturing adversaries.  Not all
adversaries can be conveniently described in terms of derivation rules.

There are other approaches that weaken the Dolev-Yao adversary
assumptions by either taking  
concrete encryption schemes into account, or at least adding new algebraic
identities to the algebra of messages. Bieber \citeyear{r:bieber90} does
not assume that the encryption scheme is a free algebra, following an idea
due to Merritt and Wolper  \citeyear{r:merritt85}. Even \etal{}
\citeyear{r:even85} analyze ping-pong protocols under RSA, taking the 
actual encryption scheme into account. The applied $\pi$-calculus of Abadi
and Fournet \citeyear{r:abadi01} permits the definition of an equational
theory over the messages exchanged between processes, weakening some
of the encryption scheme assumptions when the applied
$\pi$-calculus is used to analyze security protocols.  Since the
encryption scheme used in our framework is a simple parameter to the logic,
there is no difficulty in modifying our logic to reason about a
particular encryption scheme, and hence we can capture these approaches in
our framework. 
However, again, it seems that our approach is more flexible than these
other approaches; not all adversaries can be defined simply by starting
with a Dolev-Yao adversary and adding identities.

Recent tools for formal analysis of security protocols in the symbolic
setting have moved to a general way of describing adversaries based on
specifying the cryptosystem (including the adversary's abilities)
using an equational theory. 
This can be used to model the Dolev-Yao adversary, but also can move
beyond. 
AVISPA~\cite{r:vigano05} and ProVerif~\cite{r:abadi05} are
representative of that class of analysis tools. 
It is clear that relying on equational theories leads to the
possibility of modeling very refined adversaries, but there are
restrictions.
For instance, the equational theories used in the analysis must be
decidable, in the sense that there must exist an algorithm for
determining whether two terms in the cryptosystem are equal with
respect to the equational theory. 
Much of the recent research has focused on identifying large decidable
classes of equational theories based on some identifiable
characteristic structure of those theories
\cite{r:abadi04,r:abadi05a,r:chevalier06}. 
Clearly, we can model decidable equational theories in our setting by
simply implementing the algorithm witnessing the decidability, and in
that sense we benefit from the promising work on the subject. 
But we can also naturally support approximation algorithms for
undecidable theories in a completely transparent way, and therefore we
are not restricted in the same way. 
On the other hand, we do not seek to support automated tools, but
rather to provide an expressive framework for modeling and specifying
security protocols. 

Another class of formal approaches to security protocol analysis has
recently been proving popular, the class of approaches based on
\emph{computational cryptography}~\cite{r:goldreich01}.
These approaches take seriously the view that messages are strings of
bits, and that adversaries are efficient Turing machines, generally,
randomized polynomial-time algorithms. 
A protocol satisfies a security property in this setting if it
satisfies it in the presence of an arbitrary adversary taken from a
given class of algorithms.
Thus, this corresponds to modeling a security protocol in the presence
not of a single adversary, but rather a family of adversaries. 
Early work by Datta et al.~\citeyear{r:datta05} and more recently the
development of automated tools such as CryptoVerif~\cite{r:blanchet08}
has shown the applicability of the approach to protocol 
analysis.\footnote{An orthogonal line of research investigates the
  relationship between computational cryptography and the extent to
  which it can be soundly approximated by more symbolic
  approaches~\cite{r:backes03,r:micciancio04}. We do not address this
  question here.}
Our framework is amenable to supporting multiple adversaries via a
simple extension: rather than having a single global knowledge
algorithm, we can make the knowledge algorithm part of the local state
of the adversary---this is in fact the original and most general 
  presentation of algorithmic knowledge \cite{r:halpern94}. 
With the knowledge algorithm now part of the local state, we can model 
protocol execution in the presence of a class of adversaries, each
represented by its knowledge algorithm. 
The system generated by protocol $P$ in the presence of a class of
adversaries $\mathit{ADVs}$ is the union of the runs of $P$
executed under each adversary $A\in\mathit{ADVs}$. The initial states
of the system are the initial states of $P$ under each
adversary. 
Intuitively, each run of the system corresponds to a possible
execution of protocol $P$ under some nondeterministically chosen
adversary in $\mathit{ADVs}$. 
In the case of computational cryptography models, the knowledge
algorithms can simply represent all possible polynomial-time
algorithms. 
Such an approach to modeling systems under different adversaries is
along the lines of the model developed by Halpern and
Tuttle~\citeyear{HT}.

On a related note, the work of Abadi and Rogaway
\citeyear{r:abadi02a}, building on previous work by Bellare and Rogaway
\citeyear{r:bellare93}, compare the results obtained by a Dolev-Yao
adversary with those obtained by a more computational view of
cryptography. They show that, under various conditions, the former is
sound with respect to the latter, that is, terms that are
assumed indistinguishable in the Dolev-Yao model remain indistinguishable
under a concrete encryption scheme.  
It would be interesting to see the extent to which their analysis can
be recast in our setting, which, as we argued, can capture both the
Dolev-Yao adversary and more concrete adversaries. 

The use of a security-protocol logic based on knowledge or belief is
not new.  
Several formal logics for analysis of security protocols that
involve knowledge and belief have been introduced, going back to BAN
logic \cite{r:burrows90}, such as
\cite{r:bieber90,r:gong90,r:syverson90,r:abadi91,r:stubblebine96,r:wedel96,r:accorsi01}.  
The main problem 
with some of those approaches is that semantics of the logic (to the
extent that one is provided) is typically not tied to protocol
executions or attacks. As a result, protocols are analyzed in an
idealized form, and this idealization is itself error-prone and
difficult to formalize \cite{r:mao95}.%
\footnote{While more recent
logical approaches (e.g., \cite{r:clarke98,r:durgin03}) do not suffer from an
idealization phase and are more tied to protocol execution, they
also do not attempt to capture knowledge and belief in any general
way.} While some of these approaches have a well-defined semantics and
do not rely on idealization (e.g., \cite{r:bieber90,r:accorsi01}), they
are still restricted to (a version of) the Dolev-Yao adversary. In
contrast, our framework goes beyond Dolev-Yao, as we have seen, and
our semantics is directly tied to protocol execution.
Other approaches have notions of knowledge that can be interpreted as
a form of algorithmic knowledge (\cite{r:durgin03}, for instance), but 
the interpretation of knowledge is fixed in the semantics of the
logic. 
One limitation that our logic shares with other logics for security
protocol analysis based on multiagent systems is that we can only
reason about a fixed finite number of agents participating in the
protocol. This is in contrast to approaches such as process calculi
that can implicitly deal with an arbitrary number of agents. 

The problem of logical omniscience in logics of
knowledge is well known, and the literature describes numerous
approaches to try to circumvent it. (See \cite[Chapter 10 and
11]{r:fagin95} for an overview.) 
In the context of security, this takes the form of using different
semantics for knowledge, either by introducing \emph{hiding} operators
that hide part of the local state for the purpose of
indistinguishability 
or by using notions such as \emph{awareness} \cite{FH} to capture an
intruder's inability to decrypt \cite{r:accorsi01}.\footnote{A notion
  of algorithmic knowledge was defined by Moses \citeyear{r:moses88}
  and used by Halpern, Moses and Tuttle \citeyear{r:halpern88} to
  analyze zero-knowledge protocols. Although related to algorithmic
  knowledge as defined here, Moses' approach does not use an explicit
  algorithm. Rather, it checks whether these exists an algorithm of a
  certain class (for example, a polynomial-time algorithm) that could
  compute such knowledge.  }  
We now describe these two approaches in more detail.

The hiding approach is used in many knowledge-based frameworks as a
way to define an essentially standard semantics for knowledge not
subject to logical omniscience, at least as far as cryptography
is concerned. 
Abadi and Tuttle \citeyear{r:abadi91}, for instance, map all messages
that the agent cannot decrypt to a fixed symbol $\Box$; the semantics
of knowledge is modified so that $s$ and $s'$ are indistinguishable to
agent $i$ when the local state of agent $i$ in $s$ and $s'$ is the
same after applying the mapping described above. 
Syverson and van Oorschot \citeyear{r:syverson94} use a variant:
rather than mapping all messages that an agent cannot decrypt to the
same symbol $\Box$, they use a distinct symbol $\Box_x$ for each
distinct term $x$ of the free algebra modeling encrypted messages, and
take states containing these symbols to be indistinguishable if they
are the same up to permutation of the set of symbols $\Box_x$.  
Thus, an adversary may still do comparisons of encrypted messages
without attempting to decrypt them. 
Hutter and Schairer \citeyear{r:hutter04} use this approach in their
definition of information flow in the presence of symbolic
cryptography, and Garcia \etal{} \citeyear{r:garcia05} use it in their
definition of anonymity in the presence of symbolic
cryptography.\footnote{A variant of this apprach is developed by Cohen
  and Dam \citeyear{r:cohen05} to deal with logical omniscience in a
  first-order interpretation of BAN logic. Rather than using a symbol
  $\Box$ to model that a message is encrypted with an unknown key,
  they identify messages in different states encrypted using
an unknown key. }
This approach deals with logical omniscience for encrypted messages:
when the adversary receives a message $\msg$ encrypted with a key that he
does not know, the adversary does not know that he has $\msg$ if there
exists another state where he has received a different message $\msg'$
encrypted with a key he does not know.  
However, the adversary can still perform arbitrary computations with
the data that he does know. 
Therefore, this approach does not directly capture \emph{computational
  limitations}, something algorithmic knowledge takes into
account.

Awareness is a more syntactical approach.
Roughly speaking, the semantics for awareness can 
specify for every point a set of formulas of which an agent is
aware. For instance, an agent may be aware of a formula without being
aware of its subformulas. A general problem with awareness is
determining the set of formulas of which an agent is aware at any
point. One interpretation of algorithmic knowledge is that it
characterizes what formulas an agent is aware of: those for which the
algorithm says $\YES$. In that sense, we subsume approaches based on
awareness by providing them with an intuition. 
We should note that not every use of awareness in the security
protocol analysis literature is motivated by the desire to model more
general adversaries. 
Accorsi \etal{} \citeyear{r:accorsi01}, for instance, describe a logic
for reasoning about beliefs of agents participating in a protocol,
much in the way that BAN logic is used to reason about beliefs of
agents participating in a protocol. 
To deal with the logical omniscience problem, Accorsi \etal{} use
awareness to restrict the set of facts that an agent can believe. 
Thus, an agent may be aware of which agent sent a message if she
shares a secret with the sender of the message, and not be aware of
that fact otherwise. 
This makes the thrust of their work  different from ours.

\section{Conclusion}

We have presented a framework for 
security analysis using algorithmic knowledge. The knowledge algorithm can be 
tailored to account for both the capabilities of the adversary and the 
specifics of the protocol under consideration. 
Of course, it is always 
possible to take a security logic and extend it in an ad hoc way to 
reason about adversary with different capabilities.
Our approach has
a number of advantages over ad hoc approaches.  In particular,  it is
a general framework (we simply need to change the algorithm used by
the adversary to change its  
capabilities, or add adversaries with different capabilities), and it
permits reasoning about protocol-specific issues (for example, it can
capture situations 
such as an agent sending the bits of her key).
Another advantage of our approach is that it naturally extends to the
probabilistic 
setting. For instance, we can easily handle probabilistic 
protocols by considering multiagent systems with a probability 
distribution on the runs (see \cite{HT}). 
We can also deal with knowledge
algorithms that are probabilistic, although there are some additional
subtleties that arise, 
since the  semantics for $X_i$ given here assumes that the knowledge
algorithm is deterministic. 
In a companion paper \cite{r:halpern05c}, we extend our approach 
to deal with probabilistic algorithmic knowledge, which lets us reason
about a 
Dolev-Yao adversary that attempts to guess keys subject to a
distribution. We hope to use this approach to capture probabilistic 
adversaries of the kind studied by Lincoln \etal{} \citeyear{r:lincoln98}. 

The goal of this paper was to introduce a general framework for
handling different adversary models in a natural way, not specifically
to devise new attacks or adversary capabilities. 
In fact, finding new attacks or defining new adversaries are
difficult tasks orthogonal to the problem of using the
framework. 
One potential application for the framework would be to formalize and
compare within the same knowledge-based framework new attacks that
are introduced by the community.  We gave a concrete example of this 
with the ``guess-and-confirm'' attacks of Lowe \citeyear{r:lowe02}. 

It is fair to ask at this point what we can gain by using this framework.  
For one thing, we believe that the ability of the framework to describe
the capabilities of the adversary will make it possible to specify the
properties of security protocols more precisely.
In particular, we can now express security properties directly in
terms of knowledge, which has the advantage of matching fairly well
the informal specification of many properties, and we can give a clear
algorithmic semantics to knowledge based on the capabilities of the
adversary. More complex security properties can be expressed by
extending the logic with additional operators such as temporal
operators, extensions that are all well understood~\cite{r:fagin95}.
Of course, not all security properties can be conveniently expressed
with our logic. Properties such as computational indistinguishability
\cite{r:goldwasser84}, which is not a property of single executions
but of sets of executions, or observational equivalence
\cite{r:milner80}, which requires not only the protocol to be
analyzed but also an idealized version of the protocol that is
obviously correct, cannot be expressed directly in the logic. It
would be of interest to study the sort of extensions required to capture
those properties, and others.

Of course, it may be
the case that to prove correctness of a security protocol with respect
to certain types of adversaries (for example, polynomial-time bounded
adversaries) we will
need to appeal to techniques developed in the cryptography community.
However, we believe that it may well be possible to extend
current model-checking techniques to handle more restricted adversaries
(for example, Dolev-Yao extended with random guessing).  
This is a topic that deserves further investigation.  In any case,
having a logic where we can specify the abilities of adversaries is a
necessary prerequisite to using model-checking techniques.

\subsection*{Acknowledgments}

A preliminary version of this paper appeared in the \emph{Proceedings
of the Workshop on Formal Aspects of Security}, LNCS 2629,
pp. 115-132, 2002.

This research was inspired by discussions between the first author, Pat
Lincoln, and John Mitchell, on a wonderful hike in the Dolomites. 
We also thank Sabina Petride for useful comments.
Authors supported in part by NSF under grant 
CTC-0208535, by ONR under grants  N00014-00-1-03-41 and
N00014-01-10-511, by the DoD Multidisciplinary University Research
Initiative (MURI) program administered by the ONR under
grant N00014-01-1-0795, and by AFOSR under grant F49620-02-1-0101.

\appendix

\vspace{-50 pt}


\begin{thebibliography}{61}
\providecommand{\natexlab}[1]{#1}
\providecommand{\url}[1]{\texttt{#1}}
\expandafter\ifx\csname urlstyle\endcsname\relax
  \providecommand{\doi}[1]{doi: #1}\else
  \providecommand{\doi}{doi: \begingroup \urlstyle{rm}\Url}\fi

\bibitem[Abadi and Blanchet(2005)]{r:abadi05}
M.~Abadi and B.~Blanchet.
\newblock Analyzing security protocols with secrecy types and logic programs.
\newblock \emph{Journal of the ACM}, 52\penalty0 (1):\penalty0 102--146, 2005.

\bibitem[Abadi and Cortier(2004)]{r:abadi04}
M.~Abadi and V.~Cortier.
\newblock Deciding knowledge in security protocols under equational theories.
\newblock In \emph{Proc.~31st Colloquium on Automata, Languages, and
  Programming (ICALP'04)}, volume 3142 of \emph{Lecture Notes in Computer
  Science}, 2004.

\bibitem[Abadi and Cortier(2005)]{r:abadi05a}
M.~Abadi and V.~Cortier.
\newblock Deciding knowledge in security protocols under (many more) equational
  theories.
\newblock In \emph{Proc.~18th IEEE Computer Security Foundations Workshop
  (CSFW'05)}, pages 62--76. IEEE Computer Society Press, 2005.

\bibitem[Abadi and Fournet(2001)]{r:abadi01}
M.~Abadi and C.~Fournet.
\newblock Mobile values, new names, and secure communication.
\newblock In \emph{Proc.~28th Annual ACM Symposium on Principles of Programming
  Languages (POPL'01)}, pages 104--115, 2001.

\bibitem[Abadi and Rogaway(2002)]{r:abadi02a}
M.~Abadi and P.~Rogaway.
\newblock Reconciling two views of cryptography (the computational soundness of
  formal encryption).
\newblock \emph{Journal of Cryptology}, 15\penalty0 (2):\penalty0 103--127,
  2002.

\bibitem[Abadi and Tuttle(1991)]{r:abadi91}
M.~Abadi and M.~R. Tuttle.
\newblock A semantics for a logic of authentication.
\newblock In \emph{Proc.~10th ACM Symposium on Principles of Distributed
  Computing (PODC'91)}, pages 201--216, 1991.

\bibitem[Accorsi et~al.(2001)Accorsi, Basin, and Vigan\`{o}]{r:accorsi01}
R.~Accorsi, D.~Basin, and L.~Vigan\`{o}.
\newblock Towards an awareness-based semantics for security protocol analysis.
\newblock In Jean Goubault-Larrecq, editor, \emph{Proc.~Workshop on Logical
  Aspects of Cryptographic Protocol Verification}, volume 55.1 of
  \emph{Electronic Notes in Theoretical Computer Science}. Elsevier Science
  Publishers, 2001.

\bibitem[Backes et~al.(2003)Backes, Pfitzmann, and Waidner]{r:backes03}
M.~Backes, B.~Pfitzmann, and M.~Waidner.
\newblock A composable cryptographic library with nested operations.
\newblock In \emph{Proc.~10th ACM Conference on Computer and Communications
  Security (CCS'03)}, pages 220--230. ACM Press, 2003.

\bibitem[Baudet(2005)]{r:baudet05}
M.~Baudet.
\newblock Deciding security of protocols against off-line guessing attacks.
\newblock In \emph{Proc.~12th ACM Conference on Computer and Communications
  Security (CCS'05)}, pages 16--25. ACM Press, 2005.

\bibitem[Bellare and Rogaway(1993)]{r:bellare93}
M.~Bellare and P.~Rogaway.
\newblock Entity authentication and key distribution.
\newblock In \emph{Proc.~13th Annual International Cryptology Conference
  (CRYPTO'93)}, volume 773 of \emph{Lecture Notes in Computer Science}, pages
  232--249. Springer-Verlag, 1993.

\bibitem[Bieber(1990)]{r:bieber90}
P.~Bieber.
\newblock A logic of communication in hostile environment.
\newblock In \emph{Proc.~3rd IEEE Computer Security Foundations Workshop
  (CSFW'90)}, pages 14--22. IEEE Computer Society Press, 1990.

\bibitem[Blanchet(2008)]{r:blanchet08}
B.~Blanchet.
\newblock A computationally sound mechanized prover for security protocols.
\newblock \emph{IEEE Transactions on Dependable and Secure Computing},
  5\penalty0 (4):\penalty0 193--207, 2008.

\bibitem[Burrows et~al.(1990)Burrows, Abadi, and Needham]{r:burrows90}
M.~Burrows, M.~Abadi, and R.~Needham.
\newblock A logic of authentication.
\newblock \emph{ACM Transactions on Computer Systems}, 8\penalty0 (1):\penalty0
  18--36, 1990.

\bibitem[Chevalier and Rusinowitch(2006)]{r:chevalier06}
Y.~Chevalier and M.~Rusinowitch.
\newblock Hierarchical combination of intruder theories.
\newblock In \emph{Proc.~17th International Conference on Rewriting Techniques
  and Applications (RTA'06)}, 2006.

\bibitem[Clarke et~al.(1998)Clarke, Jha, and Marrero]{r:clarke98}
E.M. Clarke, S.~Jha, and W.~Marrero.
\newblock Using state space exploration and a natural deduction style message
  derivation engine to verify security protocols.
\newblock In \emph{Proc.~IFIP Working Conference on Programming Concepts and
  Methods (PROCOMET)}, 1998.

\bibitem[Cohen and Dam(2005)]{r:cohen05}
M.~Cohen and M.~Dam.
\newblock Logical omniscience in the semantics of {BAN} logic.
\newblock In \emph{Proc.~Workshop on Foundations of Computer Security
  (FCS'05)}, 2005.

\bibitem[Corin et~al.(2005)Corin, Doumen, and Etalle]{r:corin05}
R.~Corin, J.~Doumen, and S.~Etalle.
\newblock Analysing password protocol security against off-line dictionary
  attacks.
\newblock In \emph{Proc.~2nd International Workshop on Security Issues with
  Petri Nets and other Computational Models (WISP'04)}, volume 121 of
  \emph{Electronic Notes in Theoretical Computer Science}, pages 47--63.
  Elsevier Science Publishers, 2005.

\bibitem[Datta et~al.(2005)Datta, Derek, Mitchell, Shmatikov, and
  Turuani]{r:datta05}
A.~Datta, A.~Derek, J.~C. Mitchell, V.~Shmatikov, and M.~Turuani.
\newblock Probabilistic polynomial-time semantics for a protocol security
  logic.
\newblock In \emph{Proc.~32nd Colloquium on Automata, Languages, and
  Programming (ICALP'05)}, pages 16--29, 2005.

\bibitem[Dolev and Yao(1983)]{r:dolev83}
D.~Dolev and A.~C. Yao.
\newblock On the security of public key protocols.
\newblock \emph{IEEE Transactions on Information Theory}, 29\penalty0
  (2):\penalty0 198--208, 1983.

\bibitem[Durgin et~al.(2003)Durgin, Mitchell, and Pavlovic]{r:durgin03}
N.~A. Durgin, J.~C. Mitchell, and D.~Pavlovic.
\newblock A compositional logic for proving security properties of protocols.
\newblock \emph{Journal of Computer Security}, 11\penalty0 (4):\penalty0
  677--722, 2003.

\bibitem[Even et~al.(1985)Even, Goldreich, and Shamir]{r:even85}
S.~Even, O.~Goldreich, and A.~Shamir.
\newblock On the security of ping-pong protocols when implemented using the
  {RSA}.
\newblock In \emph{Proc.~Conference on Advances in Cryptology (CRYPTO'85)},
  volume 218 of \emph{Lecture Notes in Computer Science}, pages 58--72.
  Springer-Verlag, 1985.

\bibitem[Fagin and Halpern(1988)]{FH}
R.~Fagin and J.~Y. Halpern.
\newblock Belief, awareness, and limited reasoning.
\newblock \emph{Artificial Intelligence}, 34:\penalty0 39--76, 1988.

\bibitem[Fagin and Halpern(1994)]{FH3}
R.~Fagin and J.~Y. Halpern.
\newblock Reasoning about knowledge and probability.
\newblock \emph{Journal of the ACM}, 41\penalty0 (2):\penalty0 340--367, 1994.

\bibitem[Fagin et~al.(1995)Fagin, Halpern, Moses, and Vardi]{r:fagin95}
R.~Fagin, J.~Y. Halpern, Y.~Moses, and M.~Y. Vardi.
\newblock \emph{Reasoning about Knowledge}.
\newblock MIT Press, 1995.

\bibitem[Garcia et~al.(2005)Garcia, Hasuo, Pieters, and van Rossum]{r:garcia05}
F.~D. Garcia, I.~Hasuo, W.~Pieters, and P.~van Rossum.
\newblock Provable anonymity.
\newblock In \emph{Proc.~3rd ACM Workshop on Formal Methods in Security
  Engineering (FMSE 2005)}, pages 63--72. ACM Press, 2005.

\bibitem[Goldreich(2001)]{r:goldreich01}
O.~Goldreich.
\newblock \emph{Foundations of Cryptography: Volume 1, Basic Tools}.
\newblock Cambridge University Press, 2001.

\bibitem[Goldwasser and Micali(1984)]{r:goldwasser84}
S.~Goldwasser and S.~Micali.
\newblock Probabilistic encryption.
\newblock \emph{Journal of Computer and Systems Sciences}, 28\penalty0
  (2):\penalty0 270--299, 1984.

\bibitem[Gong et~al.(1990)Gong, Needham, and Yahalom]{r:gong90}
L.~Gong, R.~Needham, and R.~Yahalom.
\newblock Reasoning about belief in cryptographic protocols.
\newblock In \emph{Proc.~1990 IEEE Symposium on Security and Privacy}, pages
  234--248. IEEE Computer Society Press, 1990.

\bibitem[Gordon and Jeffrey(2003)]{r:gordon03}
A.~D. Gordon and A.~Jeffrey.
\newblock Authenticity by typing for security protocols.
\newblock \emph{Journal of Computer Security}, 11\penalty0 (4):\penalty0
  451--520, 2003.

\bibitem[Halpern and O'Neill(2002)]{r:halpern02a}
J.~Y. Halpern and K.~O'Neill.
\newblock Secrecy in multiagent systems.
\newblock In \emph{Proc.~15th IEEE Computer Security Foundations Workshop
  (CSFW'02)}, pages 32--46. IEEE Computer Society Press, 2002.

\bibitem[Halpern and Pucella(2003)]{r:halpern03d}
J.~Y. Halpern and R.~Pucella.
\newblock On the relationship between strand spaces and multi-agent systems.
\newblock \emph{ACM Transactions on Information and System Security},
  6\penalty0 (1):\penalty0 43--70, 2003.

\bibitem[Halpern and Pucella(2005)]{r:halpern05c}
J.~Y. Halpern and R.~Pucella.
\newblock Probabilistic algorithmic knowledge.
\newblock \emph{Logical Methods in Computer Science}, 1\penalty0 (3:1), 2005.

\bibitem[Halpern and Tuttle(1993)]{HT}
J.~Y. Halpern and M.~R. Tuttle.
\newblock Knowledge, probability, and adversaries.
\newblock \emph{Journal of the ACM}, 40\penalty0 (4):\penalty0 917--962, 1993.

\bibitem[Halpern et~al.(1988)Halpern, Moses, and Tuttle]{r:halpern88}
J.~Y. Halpern, Y.~Moses, and M.~R. Tuttle.
\newblock A knowledge-based analysis of zero knowledge.
\newblock In \emph{Proc.~20th Annual ACM Symposium on the Theory of Computing
  (STOC'88)}, pages 132--147, 1988.

\bibitem[Halpern et~al.(1994)Halpern, Moses, and Vardi]{r:halpern94}
J.~Y. Halpern, Y.~Moses, and M.~Y. Vardi.
\newblock Algorithmic knowledge.
\newblock In \emph{Proc.~5th Conference on Theoretical Aspects of Reasoning
  about Knowledge (TARK'94)}, pages 255--266. Morgan Kaufmann, 1994.

\bibitem[Hutter and Schairer(2004)]{r:hutter04}
D.~Hutter and A.~Schairer.
\newblock Possibilistic information flow control in the presence of encrypted
  communication.
\newblock In \emph{Proc.~9th European Symposium on Research in Computer
  Security (ESORICS'04)}, volume 3193 of \emph{Lecture Notes in Computer
  Science}, pages 209--224. Springer-Verlag, 2004.

\bibitem[Kripke(1963)]{r:kripke63}
S.~Kripke.
\newblock A semantical analysis of modal logic {I}: normal modal propositional
  calculi.
\newblock \emph{Zeitschrift f\"{u}r Mathematische Logik und Grundlagen der
  Mathematik}, 9:\penalty0 67--96, 1963.

\bibitem[Lincoln et~al.(1998)Lincoln, Mitchell, Mitchell, and
  Scedrov]{r:lincoln98}
P.~Lincoln, J.~C. Mitchell, M.~Mitchell, and A.~Scedrov.
\newblock A probabilistic poly-time framework for protocol analysis.
\newblock In \emph{Proc.~5th ACM Conference on Computer and Communications
  Security (CCS'98)}, pages 112--121, 1998.

\bibitem[Lowe(2002)]{r:lowe02}
G.~Lowe.
\newblock Analysing protocols subject to guessing attacks.
\newblock In \emph{Proc.~Workshop on Issues in the Theory of Security
  (WITS'02)}, 2002.

\bibitem[Lowe(1995)]{r:lowe95}
G.~Lowe.
\newblock An attack on the {Needham-Schroeder} public-key authentication
  protocol.
\newblock \emph{Information Processing Letters}, 56:\penalty0 131--133, 1995.

\bibitem[Lowe(1998)]{r:lowe98}
G.~Lowe.
\newblock Casper: A compiler for the analysis of security protocols.
\newblock \emph{Journal of Computer Security}, 6:\penalty0 53--84, 1998.

\bibitem[Mao(1995)]{r:mao95}
W.~Mao.
\newblock An augmentation of {BAN}-like logics.
\newblock In \emph{Proc.~8th IEEE Computer Security Foundations Workshop
  (CSFW'95)}, pages 44--56. IEEE Computer Society Press, 1995.

\bibitem[Meadows(1996)]{r:meadows96}
C.~Meadows.
\newblock The {NRL} protocol analyzer: An overview.
\newblock \emph{Journal of Logic Programming}, 26\penalty0 (2):\penalty0
  113--131, 1996.

\bibitem[Merritt and Wolper(1985)]{r:merritt85}
M.~Merritt and P.~Wolper.
\newblock States of knowledge in cryptographic protocols.
\newblock Unpublished manuscript, 1985.

\bibitem[Micciancio and Warinschi(2004)]{r:micciancio04}
D.~Micciancio and B.~Warinschi.
\newblock Soundness of formal encryption in the presence of active adversaries.
\newblock In \emph{Proc.~Theory of Cryptography Conference (TCC'04)}, volume
  2951 of \emph{Lecture Notes in Computer Science}, pages 133--151.
  Springer-Verlag, 2004.

\bibitem[Millen et~al.(1987)Millen, Clark, and Freedman]{r:millen87}
J.~K. Millen, S.~C. Clark, and S.~B. Freedman.
\newblock The {Interrogator}: Protocol security analysis.
\newblock \emph{IEEE Transactions on Software Engineering}, 13\penalty0
  (2):\penalty0 274--288, 1987.

\bibitem[Milner(1980)]{r:milner80}
R.~Milner.
\newblock \emph{A Calculus of Communicating Systems}.
\newblock Number~92 in Lecture Notes in Computer Science. Springer-Verlag,
  1980.

\bibitem[Mitchell et~al.(1997)Mitchell, Mitchell, and Stern]{r:mitchell97}
J.~Mitchell, M.~Mitchell, and U.~Stern.
\newblock Automated analysis of cryptographic protocols using {Mur$\varphi$}.
\newblock In \emph{Proc.~1997 IEEE Symposium on Security and Privacy}, pages
  141--151. IEEE Computer Society Press, 1997.

\bibitem[Moore(1988)]{r:moore88}
J.~H. Moore.
\newblock Protocol failures in cryptosystems.
\newblock \emph{Proceedings of the IEEE}, 76\penalty0 (5):\penalty0 594--602,
  1988.

\bibitem[Moses(1988)]{r:moses88}
Y.~Moses.
\newblock Resource-bounded knowledge.
\newblock In \emph{Proc.~2nd Conference on Theoretical Aspects of Reasoning
  about Knowledge (TARK'88)}, pages 261--276. Morgan Kaufmann, 1988.

\bibitem[Needham and Schroeder(1978)]{r:needham78}
R.~M. Needham and M.~D. Schroeder.
\newblock Using encryption for authentication in large networks of computers.
\newblock \emph{Communications of the ACM}, 21\penalty0 (12):\penalty0
  993--999, 1978.

\bibitem[Paulson(1998)]{r:paulson98}
L.~C. Paulson.
\newblock The inductive approach to verifying cryptographic protocols.
\newblock \emph{Journal of Computer Security}, 6\penalty0 (1/2):\penalty0
  85--128, 1998.

\bibitem[Pucella(2006)]{r:pucella06c}
R.~Pucella.
\newblock Deductive algorithmic knowledge.
\newblock \emph{Journal of Logic and Computation}, 16\penalty0 (2):\penalty0
  287--309, 2006.

\bibitem[Ryan and Schneider(1998)]{r:ryan98}
P.~Y.~A. Ryan and S.~A. Schneider.
\newblock An attack on a recursive authentication protocol: A cautionary tale.
\newblock \emph{Information Processing Letters}, 65\penalty0 (1):\penalty0
  7--10, 1998.

\bibitem[Stubblebine and Wright(1996)]{r:stubblebine96}
S.~Stubblebine and R.~Wright.
\newblock An authentication logic supporting synchronization, revocation, and
  recency.
\newblock In \emph{Proc.~3rd ACM Conference on Computer and Communications
  Security (CCS'96)}. ACM Press, 1996.

\bibitem[Syverson(1990)]{r:syverson90}
P.~Syverson.
\newblock A logic for the analysis of cryptographic protocols.
\newblock NRL Report 9305, Naval Research Laboratory, 1990.

\bibitem[Syverson and Cervesato(2001)]{r:syverson01}
P.~Syverson and I.~Cervesato.
\newblock The logic of authentication protocols.
\newblock In \emph{Proc.~1st International School on Foundations of Security
  Analysis and Design (FOSAD'00)}, volume 2171 of \emph{Lecture Notes in
  Computer Science}, pages 63--137, 2001.

\bibitem[Syverson and van Oorschot(1994)]{r:syverson94}
P.~F. Syverson and P.~C. van Oorschot.
\newblock On unifying some cryptographic protocol logics.
\newblock In \emph{Proc.~1994 IEEE Symposium on Security and Privacy}, pages
  14--28. IEEE Computer Society Press, 1994.

\bibitem[Thayer et~al.(1999)Thayer, Herzog, and Guttman]{r:thayer99}
F.~J. Thayer, J.~C. Herzog, and J.~D. Guttman.
\newblock Strand spaces: Proving security protocols correct.
\newblock \emph{Journal of Computer Security}, 7\penalty0 (2/3):\penalty0
  191--230, 1999.

\bibitem[Vigan\`{o}(2005)]{r:vigano05}
L.~Vigan\`{o}.
\newblock Automated security protocol analysis with the {AVISPA} tool.
\newblock In \emph{Proc.~21th Conf. Mathematical Foundations of Programming
  Semantics (MFPS'05)}, volume 155 of \emph{Electronic Notes in Theoretical
  Computer Science}, pages 61--86. Elsevier Science Publishers, 2005.

\bibitem[Wedel and Kessler(1996)]{r:wedel96}
G.~Wedel and V.~Kessler.
\newblock Formal semantics for authentication logics.
\newblock In \emph{Proc.~4th European Symposium on Research in Computer
  Security (ESORICS'96)}, volume 1146 of \emph{Lecture Notes in Computer
  Science}, pages 219--241. Springer-Verlag, 1996.

\end{thebibliography}
\end{document}